\documentclass[reprint, amsmath,amssymb,aps,prb,daraft]{revtex4-2}
\usepackage{stackengine}
\usepackage{booktabs}
\usepackage{float}

\usepackage[caption=false]{subfig}
\usepackage{graphicx}
\usepackage{bm}
\usepackage{hyperref}
\usepackage{natbib}
\usepackage[usenames,dvipsnames]{color}
\usepackage[usenames,dvipsnames,svgnames,table]{xcolor}
\usepackage{physics}
\usepackage{placeins}
\usepackage{ulem}

\usepackage{comment}
\bibpunct{\color{blue}[}{\color{blue}]}{,}{n}{}{;}
\hypersetup{
  colorlinks,
  citecolor=blue,
  linkcolor=blue,
  urlcolor=blue
  }
  \renewcommand{\vec}[1]{{\boldsymbol #1}}
\newcommand{\jk}[1]{\textcolor{blue}{#1}}
\newcommand{\rh}[1]{\textcolor{cyan}{#1}}

\newlength{\mysize}

\begin{document}
\preprint{APS/123-QED}
\title{Josephson transistor from the superconducting diode effect in domain wall and skyrmion magnetic racetracks}
\author{Richard Hess}
\author{Henry F. Legg}
\author{Daniel Loss}
\author{Jelena Klinovaja}%
\affiliation{%
 Department of Physics, University of Basel, Klingelbergstrasse 82, CH-4056 Basel, Switzerland }%
 \date{\today}
 \begin{abstract}
In superconductors, the combination of broken time-reversal and broken inversion symmetries can result in a critical current being dependent on the direction of current flow. This phenomenon is known as superconducting diode effect (SDE) and has great potential for applications in future low-temperature electronics. Here, we investigate how magnetic textures such as domain walls or skyrmions on a racetrack can be used to control the SDE in a Josephson junction and how the SDE can be used as a low-temperature read-out of the data in racetrack memory devices. First, we consider a two-dimensional electron gas (2DEG) with strong spin-orbit-interaction (SOI) coupled to a magnetic racetrack, which forms the weak-link in a Josephson junction. In this setup, the exchange coupling between the magnetic texture and the itinerant electrons in the 2DEG breaks time-reversal symmetry and enables the SDE. When a magnetic texture, such as a domain wall or skyrmion enters the Josephson junction, the local exchange field within the junction is changed and, consequently, the strength of the SDE is altered. In particular, depending on the position and form of the magnetic texture, moving the magnetic texture can cause the SDE coefficient to change its sign, enabling a Josephson transistor effect with potentially fast switching frequencies. Further, we find that the SDE is enhanced if the junction length-scales are comparable with the length-scale of the magnetic texture. Furthermore, we show that, under certain circumstances, the symmetry breaking provided by particular magnetic textures, such as skyrmions, can lead to an SDE even in the absence of Rashba SOI in the 2DEG. Our results provide a basis for new forms of readout in low-temperature memory devices as well as demonstrating how a Josephson transistor effect can be achieved even in the absence of an external magnetic field and intrinsic Rashba SOI.
\end{abstract}

\maketitle

\section{Introduction}

 \begin{figure}[!t]
\subfloat{\label{figSchematicsJJWithSkyrmion}\stackinset{l}{-0.00in}{t}{-0.0in}{\color{white}(a)\color{black}}{\includegraphics[width=1\columnwidth]{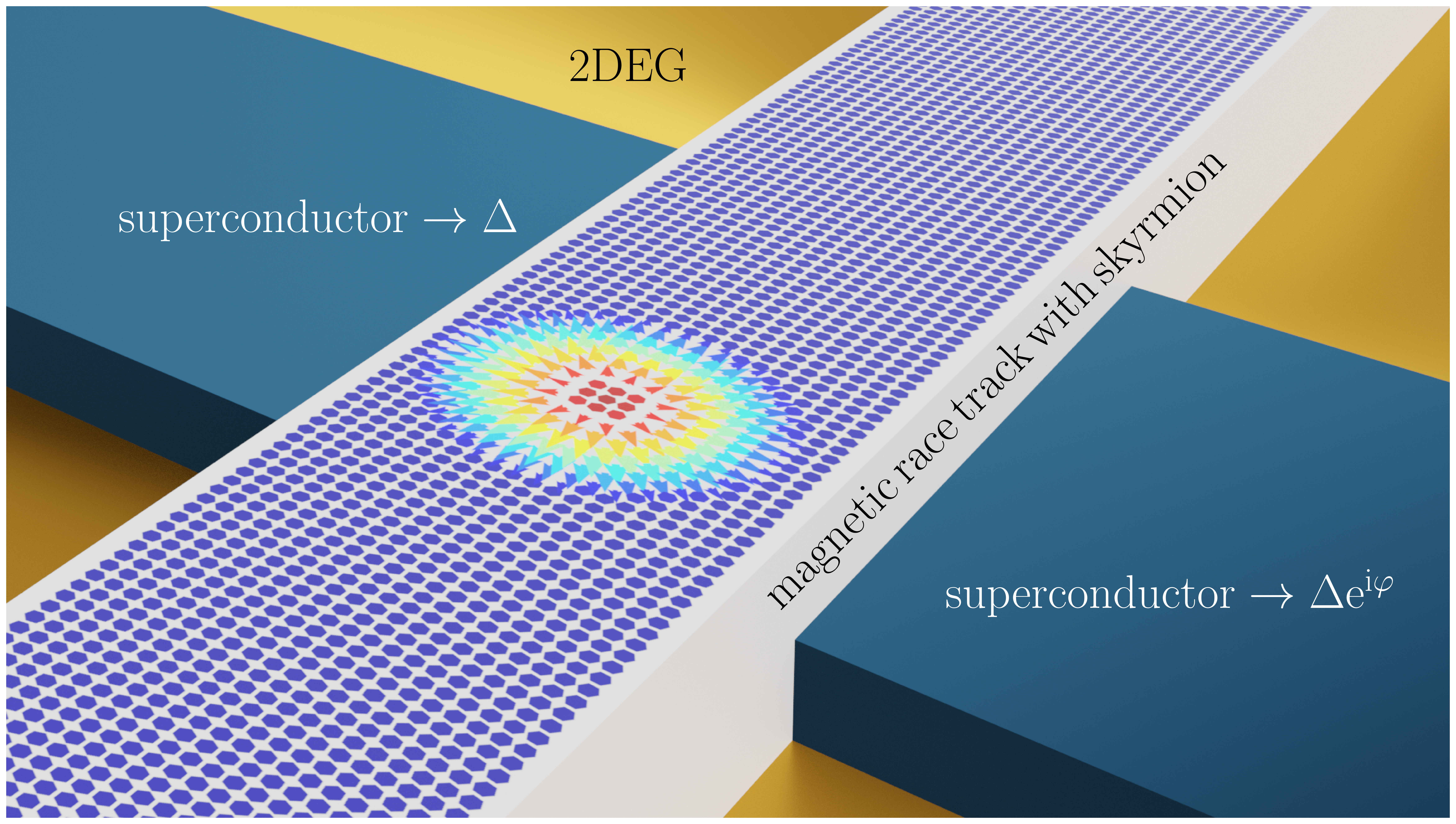}}
}
 \\
\subfloat{\label{figSchematicAC_CurrentProbe}\stackinset{l}{-0.125in}{t}{0.in}{(b)}{\stackinset{l}{-0.125in}{t}{0.75in}{(c)}{\stackinset{l}{-0.125in}{t}{1.5in}{(d)}{\includegraphics[width=0.9\columnwidth]{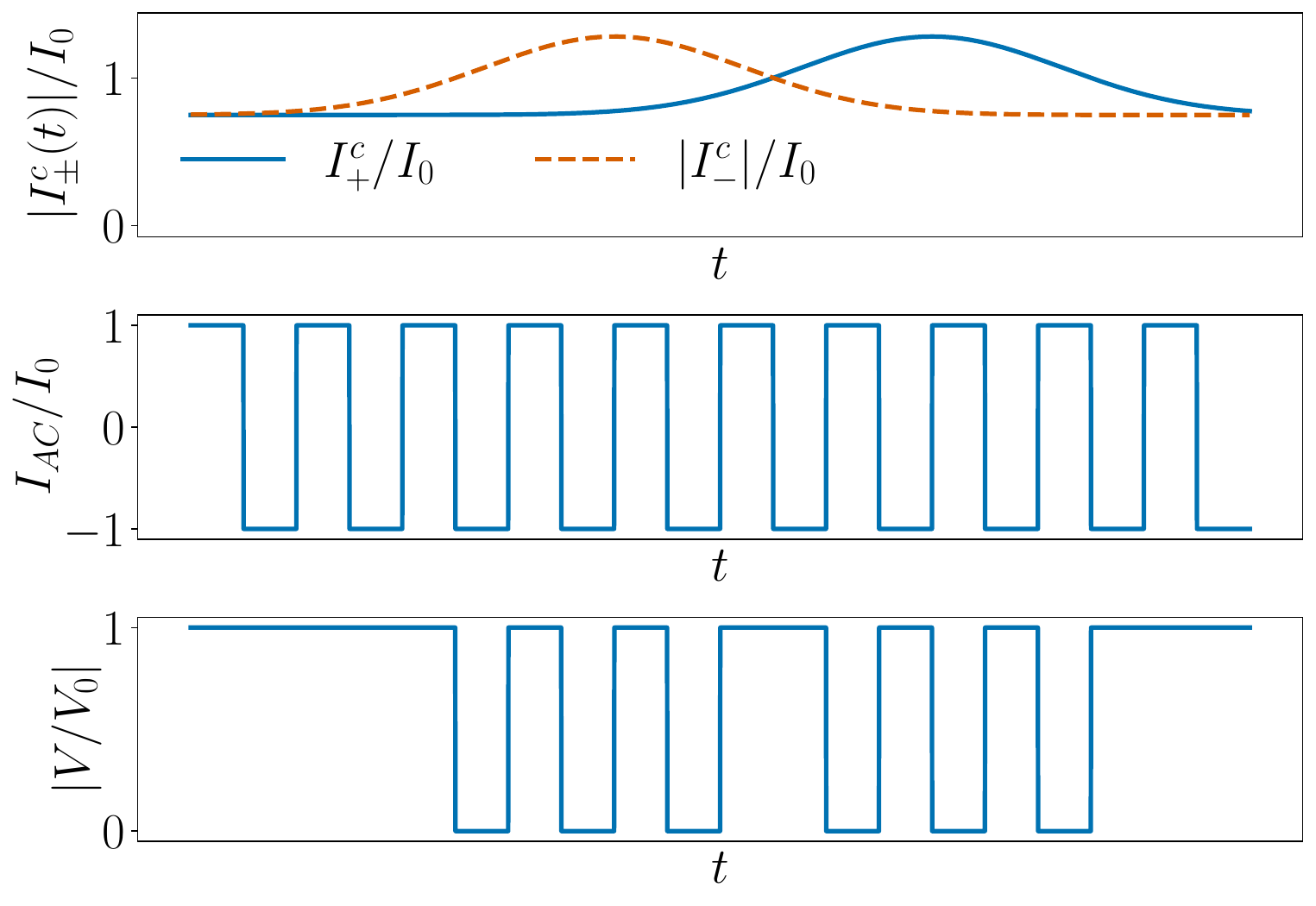}}}}
}
\subfloat{\label{figSchemticsCritCurrent}}
\subfloat{\label{figSchemticsVoltageResponse}}
\caption{ \textit{Schematic setup of a Josephson junction sandwiching a magnetic racetrack}: (a) The superconductors, described by the order parameters $\Delta$ and $\Delta e^{i\varphi}$, so that the superconducting phase difference is given by $\varphi$, are shown in blue, while the racetrack is shown in white. A skyrmion (multi-colored points and arrows) is embedded in a ferromagnetic-out-of-plane background (dark blue dots).  (b)  The positive [negative] critical super current $I_{+}^c$ [$I_{-}^c$] changes as a function of time $t$  when a magnetic texture like a domain wall or skyrmion passes through the system. 
(c) If an alternating current $I_{\rm AC}$  with amplitude $I_0$ 
is driven through the junction as a function of the time $t$, then the change of critical currents and SDE becomes visible in (d) the voltage response signal: The voltage drops if the alternating current is smaller than the critical supercurrents. Here, we normalized the voltage response signal by the voltage strength $V_0$, which is measured if $I_{\rm AC}$ is larger than the critical supercurrent in a given direction.}
 \label{figSchematics}
\end{figure}

One of the building blocks of semiconductor technology is the diode \cite{Braun1875Ueber} that, due to inversion symmetry breaking, is characterized by different values of resistances for currents flowing in opposite directions and is the basic element required to build a transistor. A similar effect, the so-called \textit{superconducting diode effect} (SDE) appears in superconductors and hybrid superconductor-semiconductor devices with broken time reversal and inversion symmetry \cite{Rikken1997Observation,Rikken2001Electrical, Wakatsuki2017Nonreciprocal, Hoshino2018Nonreciprocal, Yasuda2019Nonreciprocal, Noah2022Supercurrent, Qin2017Superconductivity, Souto2022Josephson, He2022Nonreciprocal, Wu2022Field, Lin2022ZeroField, nadeem2023superconducting, Daido2022Intrinsic,Kononov2020OneDimensional, mazur2022gatetunable, hou2023ubiquitous, Illic2022Theory,Baumgartner2022Effect, Fominov2022Asymmetric, Margarita2022Universal, Pal2022Josephson, Bauriedl2022Supercurrent,Banerjee2022Control,sinner2023superconducting}. In particular, the SDE results in critical currents that are dependent on the direction of current flow. As a consequence, for a range of currents, the SDE results in a zero resistance state in one direction, but finite Ohmic resistance in the opposite direction. 

The SDE can appear both in bulk superconductors and in Josephson junctions. Some prominent platforms that result in the SDE are artificial superconducting superlattices that lack an inversion symmetry center \cite{Fuyuki2020Observation} and two-dimensional electron gases (2DEG) with strong spin orbit interaction (SOI) brought into proximity with a superconductor \cite{Baumgartner2022Supercurrent,lotfizadeh2023superconducting}. In such setups, time-reversal symmetry is broken by an external magnetic field. The direction of the magnetic field is crucial and should couple to the inversion symmetry breaking term in the Hamiltonian \cite{Noah2022Supercurrent}, i.e. SOI, in order for a finite SDE to occur. 

Recently very large SC diode efficiencies have been achieved, opening the pathway to significant potential technological applications \cite{Wu2022FieldFree,ciaccia2023gate}. In addition to an element within future low temperature electronics, it has been proposed that the SDE can be used as a method to detect SOI strength in the presence of a superconductor \cite{Legg2022Superconducting,lotfizadeh2023superconducting} and as a measure of whether a system has entered a topological phase for example in Rashba or TI nanowires \cite{Legg2022Superconducting,legg2023parity}. 

Another promising future technology is racetrack memory devices~\cite{Parkin2008Magnetic, Ryu2013chiral, Parkin2015Memory}. The basic idea of a racetrack memory is to store information using magnetic domains in a thin quasi-one-dimensional racetrack. One advantage of racetrack memory is that the device architecture does not rely on moving parts unlike, for instance, a hard disk drive. In a racetrack memory device, currents push magnetic domains along the racetrack \cite{Yamaguchi2004RealSpace}, which can also enable a much faster read out of the stored data compared to other storage devices. In a standard racetrack setup, the magnetic domains are separated by finite size domain walls, within which the magnetization direction smoothly changes. Alternatively, however, these magnetic domains can be replaced by other spin textures such as magnetic skyrmions \cite{Yu2012Skyrmion, Sampaio2013Nucleation,Tomasello2014Strategy, Mueller2017Magnetic}.

The low operating temperature of quantum computers, for instance, has recently resulted in significantly increased interest in electronic elements, both classical and quantum, that work at low temperatures. In particular, these low temperatures enable building basic electronic devices such as transistors and read/write components from superconductors. The use of superconductors in low-temperature electronics also opens up the potential for novel and potentially faster computational devices than room temperature equivalents \cite{Wu2022FieldFree}.

In this paper, we consider the interplay of the SDE and magnetic textures on a racetrack. In particular, we show that the SDE can be controlled by magnetic domain walls or skyrmions moving on a racetrack that is sandwiched by a Josephson junction. The control of the SDE by the magnetic texture provides the basis for new low-temperature electronic components such as Josephson transistors as well as for new mechanisms for low-temperature read-out of data in racetrack memory devices. The schematic setup and functionality is shown in Fig.~\ref{figSchematics}:  A magnetic racetrack  (white) is sandwiched by two superconductors (blue) placed on top of a substrate (yellow). Many proposed magnetic racetrack materials have itinerant electrons with strong SOI, however, if the racetrack material is insulating, it can be further coupled to a 2DEG with Rashba SOI to produce an SDE. In Fig.~\ref{figSchematics}, the dots and arrows on top of the racetrack indicate the local magnetization. Here, for example, a magnetic skyrmion is embedded in a ferromagnetic background. The critical currents associated with the Josephson junction are altered when a domain wall  or skyrmion passes through the junction and therefore the critical currents also vary as a function of time, as shown schematically in Fig.~\ref{figSchemticsCritCurrent}. If an alternating current is driven through the Josephson junction, see Fig.~\ref{figSchematicAC_CurrentProbe}, then a finite voltage occurs only when the magnitude of the current in a given direction is larger than the magnitude of the critical current in that direction, see Fig.~\ref{figSchemticsVoltageResponse}. This change of the voltage signal can serve as an indicator whether a magnetic texture like a domain wall or a skyrmion passes the junction. The fact that the SDE is strongly dependent on the position of the texture can also enable a Josephson transistor effect. 
Furthermore, importantly, we find that the length-scales such as the ratio between the Fermi wave length in the two-dimensional electron gas and the skyrmion size strongly influence the diode efficiency as a function of the position of the magnetic texture. 
 
In the second part of this paper we consider Josephson junctions hosting racetracks with arbitrary smoothly spatially varying magnetic textures but now in the absence of Rashba SOI. Most proposals for the SDE in Josephson junctions rely on the presence of Rashba SOI in the 2DEG. However, it is known that non-uniform magnetic textures can map to a combination of a uniform exchange coupling field and some effective SOI \cite{Braunecker2010SpinSelective, Choy2011Majorana, Martin2012Majorana, Perge2013Proposal, Nakosai2013TwodDimensional, Chen2015Majorana, Yang2016Majorana, Gangadharaiah2011Majorana, Klinovaja2011Transition, Klinovaja2013TopologicalSuper, Klinovaja2013Giant,  Braunecker2013Interplay, Vazifeh2013Self, Hsu2015Antiferromagnetic, Rex2019Majorana, Garnier2019Topological, Mascot2021Atomic, Diaz2021Majorana}.  A helical spin chain, for example, maps to a ferromagnetic chain with Rashba SOI \cite{Braunecker2010SpinSelective, Hess2022Prevalence}. As such, we show that an intrinsic Rashba SOI in the 2DEG is not a necessary ingredient for the SDE or Josephson transistor effect in our setup and spatially-varying magnetic textures within the Josephson junction by themselves can be sufficient to result in the SDE. 

This paper is organized as follows: First, in Sec.~\ref{Sec:Model}, we define a simple model describing a quasi-two-dimensional electron gas with exchange coupling to the magnetization of a racetrack and which is sandwiched by two superconductors forming a Josephson junction. In addition, we describe details about the how we numerically perform calculations of critical current.  Second, in Sec.~\ref{Sec:SDEForFerromagneticSystems}, we analyze the SDE  for a ferromagnetic texture as a function of the chemical potential and the exchange coupling along the junction with a focus on sign changes of the diode efficiency. 
Next, in Sec.~\ref{Sec:ScDiodeEffectOnARaceTrack},  we discuss the SDE for domain walls and skyrmions moving on the racetrack. We classify smooth magnetic textures in Sec.~\ref{Sec_SDE_Without_ExplicitSOI} and predict which texture-class can mediate a SDE without the need of explicit Rashba SOI in the 2DEG. Finally, we discuss the experimental realization and implications in Sec.~\ref{Sec:Conclusion}. In Appendix~\ref{Sec:AppGaugeTrafo}, we present details on the gauge transformation used in Sec.~\ref{Sec_SDE_Without_ExplicitSOI} and, in Appendix~\ref{Sec:AppMagTexAndSDE}, we choose three random examples from different classes of magnetic textures and calculate the corresponding SC diode efficiencies: these numerical results confirm the predictions made in Sec.~\ref{Sec_SDE_Without_ExplicitSOI}. Last, in Appendix~\ref{App:PosDomainWallSkyrmion}, we clarify  notations used throughout the paper.

\section{Model \label{Sec:Model}}
We utilize an effective two-dimensional (2D) tight binding model to describe  a Josephson junction with a normal section within which the exchange coupling to the magnetic racetrack occurs. The kinetic contribution $H_{\mathrm{kin}}$ to the full Hamiltonian is given by
\begin{align}
H_{\mathrm{kin}}=-\sum_{\langle \mathbf{n},\mathbf{m}\rangle,\nu} t c_{ \mathbf{n},\nu}^{\dagger}c_{\mathbf{m},\nu}+ \sum_{ \mathbf{n},\nu}(4t-\mu_{\mathbf{n}}) c_{ \mathbf{n},\nu}^{\dagger}c_{ \mathbf{n},\nu}  \label{eq:HKin},
\end{align}
where $t=\hbar^2/(2m_{\rm{eff}}a^2)$ and $\mu_{\mathbf{n}}$ denote the hopping amplitude and the position dependent potential, respectively. Here, $m_{\rm{eff}}$ is the effective mass of the itinerant electrons and $a$ is the lattice constant.
Moreover, $\mathbf{n}=(n_x,n_y) \, [ {\rm{or }}\,\, \mathbf{m}=(m_x,m_y)]$ denote the coordinate of a lattice site and  $\nu$ denotes the spin $\uparrow,\downarrow$ along the quantization axis, so that $c_{\mathbf{n},\nu}^{\dagger}$ ($c_{\mathbf{n},\nu}$) creates (annihilates) an electron with spin $\nu$ at the site $\mathbf{n}$. Here, the first sum runs over nearest neighbour sites as indicated by the notation $\langle \mathbf{n},\mathbf{m}\rangle$.
The superconducting pairing potential is modelled via
\begin{align}
H_{\mathrm{sc}}=\sum_{\mathbf{n}}\left(\Delta_{\mathbf{n}}c_{\mathbf{n},\uparrow}^{\dagger} c_{\mathbf{n},\downarrow}^{\dagger}  +\Delta_{\mathbf{n}}^*c_{\mathbf{n},\downarrow} c_{{\mathbf{n}},\uparrow} \right),
\end{align}
where $\Delta_{\mathbf{n}}$ denotes the local superconducting pairing potential at site $\mathbf{n}$. 
The coupling between itinerant electrons and the magnetization texture is described by 
\begin{align}
H_{J}=\sum_{\mathbf{n},\nu,\nu'} J_{\mathbf{n}} \left[\boldsymbol{\sigma}\cdot\mathbf{S_n} \right]_{\nu,\nu'} c_{\mathbf{n},\nu}^{\dagger}c_{\mathbf{n},\nu'},
\end{align}
where $J_{\mathbf{n}}$ describes the exchange coupling strength between the spin $\boldsymbol{\sigma}$ of the itinerant electrons and the local magnetic moments 
\begin{align}
\mathbf{S}_{\mathbf{n}}=\begin{pmatrix}
\cos[\vartheta(\mathbf{n})]\sin[\Phi(\mathbf{n})] \\
\sin[\vartheta(\mathbf{n})]\sin[\Phi(\mathbf{n})] \\
\cos[\Phi(\mathbf{n})]
\end{pmatrix}, \label{eq:MagnetizationVector}
\end{align}
which we treat classically.
Here, $\Phi(\mathbf{n})$ and $\vartheta(\mathbf{n})$ are the polar and azimuthal angles, respectively, at the lattice site $\mathbf{n}$.
Next, we account for Rashba SOI via
\begin{align}
&H_{{\mathrm{so}}}=\alpha_l\sum_{n_x,n_y}\Bigg[c_{\downarrow,n_x-1,n_y}^{\dagger}c_{\uparrow,n_x,n_y}-c_{\downarrow,n_x+1,n_y}^{\dagger}c_{\uparrow,n_x,n_y}
\nonumber \\
&+ i\left(c_{\downarrow,n_x,n_y-1}^{\dagger}c_{\uparrow,n_x,n_y}-c_{\downarrow,n_x,n_y+1}^{\dagger}c_{\uparrow,n_x,n_y}\right) + \text{H.c.}
\Bigg], \label{eq:HamSOI}
\end{align}
with $\alpha_l=\alpha/(2a)$ the finite-difference version of the Rashba SOI strength $\alpha$ 
\cite{Sato2009NonAbelian, Dmytruk2020Pinning}.   The full Hamiltonian is then given by 
\begin{align}
H=H_{\mathrm{kin}}+H_{\mathrm{sc}}+H_{J}+H_{\mathrm{so}}. \label{eq:FullTightBindingHamiltonian}
\end{align}

We define the parameter profiles as follows: The local superconducting pairing potential is described by
\begin{align}
\Delta_{\mathbf{n}}=\Delta \Theta(N_L-n_x)+ \Delta e^{i\varphi}\Theta(n_x-N_R),
\end{align}
where $N_L$ ($N_R$) defines the position of the left (right) interface between superconducting and normal region, so that the width of the junction in terms of lattice sites is set by $N_J=N_R-N_L$. The angle $\varphi \in [0,2\pi)$ is the phase difference between left and right superconductor.  Here, we used the Heaviside function $\Theta$ with the particular definition $\Theta(0)=1$. 
Second, we define 
\begin{align}
J_{\mathbf{n}}= J [\Theta(N_R-n_x)-\Theta(N_L-n_x)],
\end{align}  
so that the effective magnetization is only non-zero inside the junction and with a uniform exchange coupling strength, $J$, to the spins of the itinerant electrons. Finally, we define the local potential
\begin{align}
\mu_{\mathbf{n}}=\mu+\gamma(\delta_{n_x,N_L}+\delta_{n_y,N_R}),
\end{align}
where we accounted for tunnel barriers at the superconductor normal (SN) interface. The symbols $\mu$ and $\gamma$ denote the chemical potential and the barrier strength, while $\delta_{n,m}$ denotes the Kronecker delta. 

\begin{figure*}[t]
\subfloat{\label{figDomainWallConfiguration1}\stackinset{l}{-0.00in}{t}{0.1in}{}{\includegraphics[width=1\textwidth]{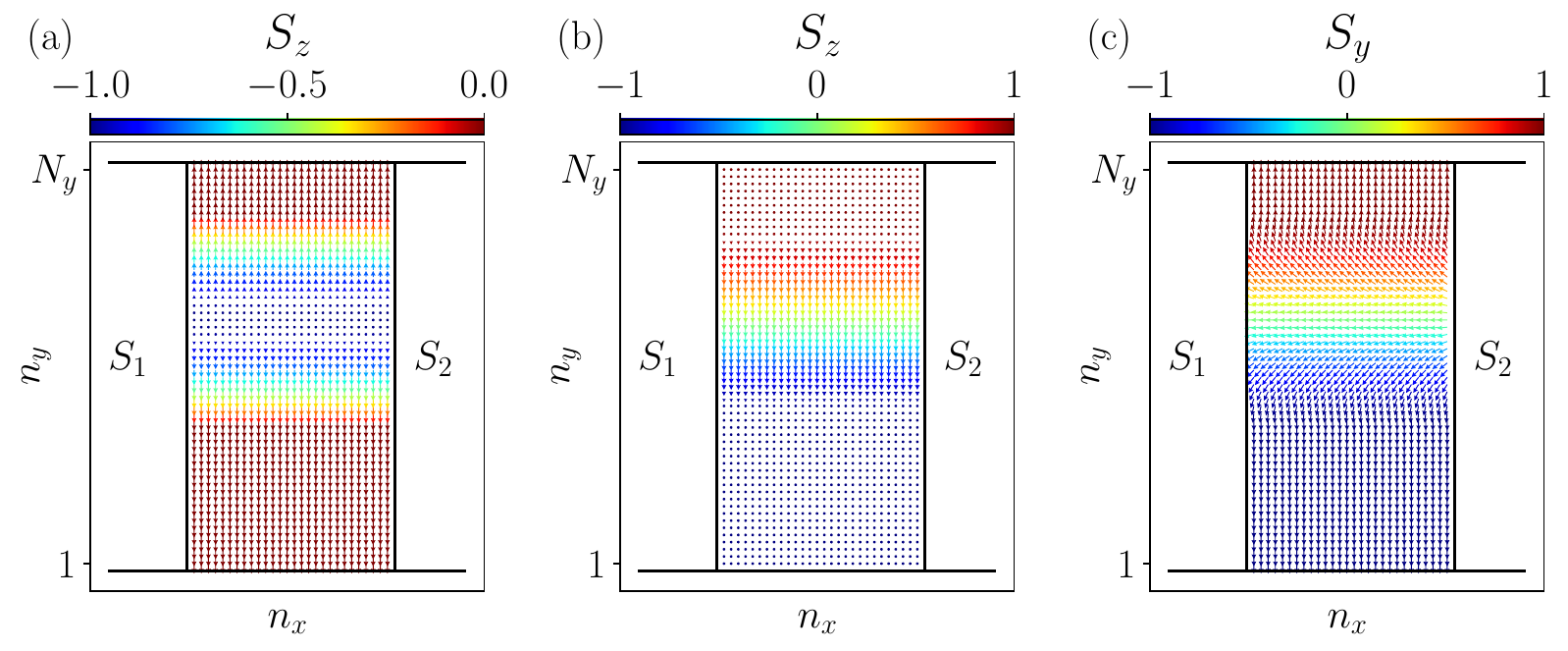}}
}
\subfloat{\label{figDomainWallConfiguration2}}
\subfloat{\label{figDomainWallConfiguration3}}
\caption{ \textit{Schematic representation of the domain wall profiles}:  Here, $S_1$ and $S_2$ denote the left and right superconductor, the magnetization is only non-zero in between the superconducting regions. In particular, the arrows and the colorbars indicate the in-plane and out-of-plane orientation of the magnetic texture, except in panel (c), where the colorbar shows the $S_y$ component, since the $S_z$ component is zero. The domain wall described by the angles (a)  $\vartheta_{dw,1}$ and $\Phi_{dw,1}$;  (b) $\vartheta_{dw,2}$ and $\Phi_{dw,2}$; (c) $\vartheta_{dw,3}$ and $\Phi_{dw,3}$. We note that only the components pointing in $y$-direction contribute to the superconducting diode effect in our case. \label{figDomainWallConfiguration}
 Parameters: $L_y=2L_x=140$ nm, $a=2.5$ nm, and $\lambda_{dw}a=70$ nm.
 }
\end{figure*}

\subsection{Magnetization profiles}

Here, we define the different types of 
magnetization profiles that will be utilized throughout the paper. Namely, various types of domains walls and skyrmions.
\subsubsection{Domain walls}
We will analyze three different profiles of magnetic domain walls. First, we  consider a domain wall as described by $\vartheta_{dw,1}(n_y)=\frac{\pi}{2}$  and 
\begin{align}
\Phi_{dw,1}(n_y)= \begin{cases} \frac{\pi}{2}, & n_y \leq n_{dw},\\ 
\frac{ \pi(n_y-n_{dw})}{\lambda_{dw}} +\frac{\pi}{2},& n_{dw} <n_y<\lambda_{dw}+n_{dw}, \\
 \frac{3\pi}{2},&  n_y \geq \lambda_{dw} +n_{dw}, \label{eq:DWAngleInPLaneMagnetization}
\end{cases}
\end{align}
where $n_{dw}$ and $\lambda_{dw}$ determine the $y$-coordinate of the first site of the domain wall and its size, respectively. This choice of $\Phi(n_y)$ models a  magnetization out of plane at the center, $n_c=n_{dw}+\lambda_{dw}/2$, of the domain wall and  it describes a magnetization parallel or anti-parallel to the $y$-direction for large distances ($|n_y-n_{c}|>\lambda_{dw}/2$) away  from the domain wall center, see Fig.~\ref{figDomainWallConfiguration1}. The substitution $\Phi_{dw,2}(n_y)\Rightarrow \Phi_{dw,1}(n_y)\pm\pi/2$ leads to a magnetization aligned out of plane for large distances away from the center of the domain wall, which is the second configuration analyzed in this paper. Finally, we define a domain wall with $\Phi_{dw,3}(n_y)=\frac{\pi}{2}$ and $\vartheta_{dw,3}(n_y)=\Phi_{dw,1}(n_y)$ modelling an in-plane magnetization aligned along the $y$-direction for large distances (larger than $\lambda_{dw}/2$) away from the domain wall center and aligned in $x$-direction at $n_c$, see Fig.~\ref{figDomainWallConfiguration1}.

\subsubsection{Skyrmions \label{Sec_SubSec_Skyrmions}}
In addition to magnetic domain walls, we consider Néel \cite{Heinze2011Spontaneous} and Bloch \cite{Yu2010Real} type skyrmions.  The Néel skyrmion is described by a polar angle of the form
\begin{align}
\Phi_{ns}(\mathbf{n})=\begin{cases} \pi & \text{if } r>\lambda_{s}, \\
\pi r/\lambda_{s} &\text{otherwise},
\end{cases}
\label{eq:DefSkyrmionTheta}
\end{align}
where $n_{u,s}$ with $u\in\lbrace x,y \rbrace$ denotes the $x$- and $y$-coordinate of the center of the skyrmion and $\lambda_s$ sets the length-scale of the skyrmion. Moreover, we introduced the quantity $r=\sqrt{(n_x-n_{x,s})^2+(n_y-n_{y,s})^2}$ measuring the distance from the center of the skyrmion and the vector
\begin{align}
\mathbf{r}=\begin{pmatrix}
n_x-n_{x,s} \\
n_y-n_{y,s} 
\end{pmatrix} =
\begin{pmatrix} 
r\cos[\vartheta_{ns}(n_x,n_y)] \\
r\sin[\vartheta_{ns}(n_x,n_y)] 
\end{pmatrix},
\end{align} 
which defines the azimuthal angle $\vartheta_{ns}(\mathbf{n})$ measured from the position of the skyrmion.
The angle of the Bloch skyrmion is related to the Néel skyrmion angle via $\vartheta_{bs}\rightarrow\vartheta_{ns}-\frac{\pi}{2}$.

\subsection{Calculation of the current \label{App:CurrCalc}}
In this subsection, we present the details on the calculations of the supercurrents. The computation is mainly  based on the Heisenberg equation of motion \cite{Furusaki1994DcJosephson,Rodero1994Microscopic, Yeyati1995SelfConsistent}, which, in general, supports the computation of local currents. Here, however, we are mainly interested in the total current passing in $x$-direction through the system.  This total current is conserved inside the junction and therefore the total current does not depend on the $x$-coordinate, as long as it is located in the junction.  In contrast, the calculation of the current inside the superconductor requires a self-consistent calculation of the superconducting order parameter to ensure current conservation, this however is not considered here, therefore we follow the calculations presented in Refs. \cite{Ostroukh2016TwoDimensional,Zuo2017Supercurrent, Vries2018Superconducting, Baumgartner2022Transport, Himmler2022Supercurrents}. 

\begin{figure}[t]
 \centering
\subfloat{\label{figSchematicsLattice}\stackinset{l}{-0.00in}{t}{-0.0in}{}{\includegraphics[width=0.75\columnwidth]{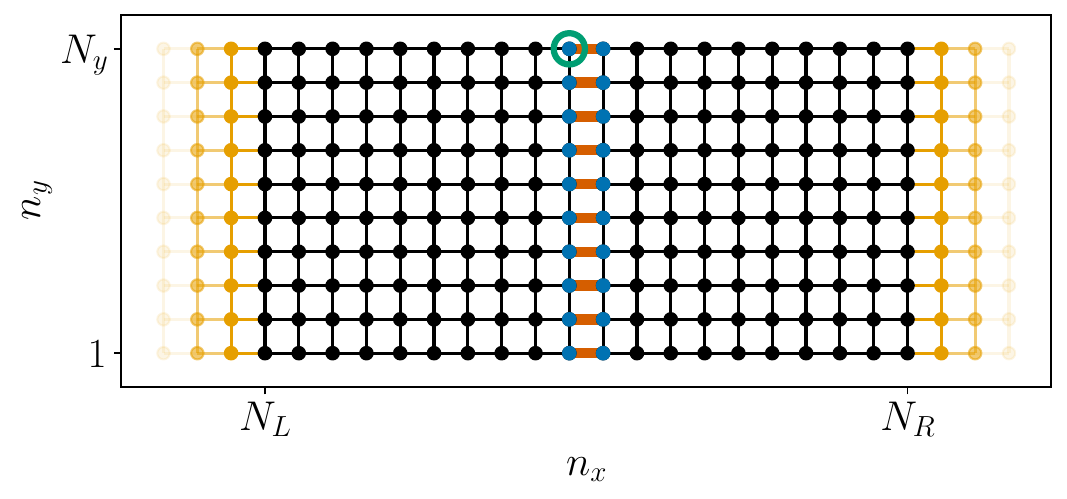}}
}
\caption{ \textit{Schematic illustration of the implemented tight binding model}: The normal region is represented by the black and blue sites, while the superconducting leads are represented by the yellow sites. The total current in $x$-direction is calculated by summing up the local currents on the red bonds between the blue colored sites. We note that the number of sites shown here does not match with the number of sites used in the actual calculations. }
 \label{figSchematicsLattice2}
\end{figure}

The local current between two adjacent lattice  sites $\mathbf{n}$ and $\mathbf{m}$ is given by
\begin{align}
I_{\mathbf{n},\mathbf{m}}=2\frac{ek_BT}{\hbar}\sum_{n=0}^{\infty}\text{Im}&\left\lbrace\text{Tr}\left[H_{\mathbf{n},\mathbf{m}}G_{\mathbf{m},\mathbf{n}}(i\omega_n)\right.\right. \nonumber \\
&\left. \left. -H_{\mathbf{m},\mathbf{n}}G_{\mathbf{n},\mathbf{m}}(i\omega_n)\right]\right\rbrace, \label{Eq:CurrentfromGreensFunc}
\end{align}
where $k_B$ and $T$ denote the Boltzmann constant and the temperature of the system, respectively \cite{Ostroukh2016TwoDimensional, Zuo2017Supercurrent, Vries2018Superconducting, Baumgartner2022Transport, Himmler2022Supercurrents}. Moreover, $H_{\mathbf{n},\mathbf{m}}$ [$G_{\mathbf{n},\mathbf{m}}$] is the submatrix of the Hamiltonian [Green's function] that connects the sites $\mathbf{n}$ and $\mathbf{m}$. In addition, $\omega_n=(2n+1)\pi k_BT$ are the fermionic Matsubara frequencies and the corresponding summation over $n$ can be carried out numerically due to a fast convergence, which enables a truncation of the sum when the required accuracy is reached. Next, we define the total current $I_x= \sum_{\mathbf{n} \in \Upsilon } I_{\mathbf{n},\mathbf{n}+\mathbf{e}_x}$  in $x$-direction as the sum of all local currents through a cross section in $y$-direction in between two adjacent columns of sites. Here, $\mathbf{e}_x$ denotes the unit vector in $x$-direction. For example, the total current would be the sum of the all local currents flowing through the red bonds connecting the blue colored sites in Fig.~\ref{figSchematicsLattice2}. Here, we denote the set of sites to the left of the cross section as $\Upsilon$, which corresponds to the left blue colored sites in Fig.~\ref{figSchematicsLattice2}.

Our numerical calculations are based on the Python package Kwant \cite{Groth2014Kwant}. 
In most of the work, we discretize the normal region (black and blue colored sites in Fig.~\ref{figSchematicsLattice2}) and attach superconducting semi-infinite leads to the left and right (yellow colored sites in Fig.~\ref{figSchematicsLattice2}). These leads do not have any exchange field, $J=0$. The length of the junction $L_y=N_y a$ is set by the number of sites along the $y$-direction. Although the actual width of the system with leads is infinite along the $x$-direction, we will refer to system width as $L_x=N_J a$, i.e., the number of sites describing the width of the junction.
Additionally, within the Kwant software, we attach virtual self-energy leads at the blue sites next to the cross section in order to compute the Green's function locally on these sites. A gauge transformation enables us to account for the superconducting phase difference via a complex phase added to the sub-matrix $H_{LR}$ ($H_{RL}$). In order to improve the code efficiency we follow Ref. \citenum{Ostroukh2016TwoDimensional}. In particular, we calculate the zero-phase Green's function for a given Matsubara frequency and afterwards we exploit the Dyson series to obtain the Green's function for finite phase difference $\varphi$. We exploit the same scheme to obtain the local density of states (LDOS), however, this time we calculate the retarded Green's function $G(\omega)$ for a normal frequency $\omega$ [instead for a Matsubara frequency $G(i\omega_n)$]
\begin{align}
\rho(\omega, \mathbf{n})=-\frac{1}{\pi}\text{Tr}_{{\mathbf{n}}}\left\lbrace\text{Im}\left[G(\omega+ i\kappa)\right]\right\rbrace.
\end{align}
Here, $\text{Tr}_{{\mathbf{n}}}$  indicates that we perform a partial trace, such that we only account for the Green's function submatrix associated with the site $\mathbf{n}$.  Moreover, the parameter $\kappa$ accounts for broadening, e.g., due to temperature. In this paper, we focus on the LDOS at the end of the cross section, in particular at the green encircled site.

In addition to the method described above, we implemented a separate tight-binding model, where we replace the superconducting leads by finite-size superconducting regions. We use this model to check our results for the LDOS and current. In these finite-size systems we consider the length of the superconducting regions to be larger than the superconducting coherence length $\xi$, meaning that $N_L a=(N_x-N_R) a>\xi $ holds, where $N_x$ denotes this time the total number of sites in $x$-direction including the superconducting regions. The finite size of the system enables extraction of all eigenvalues and eigenvectors, so that we can compare the sub-gap eigenvalues with the peaks found in the LDOS calculation. Moreover, we compared selected results for the current obtained from the Heisenberg equation of motion, as described above, with the current obtained from the free energy, which is given by
\begin{align}
I_x(\varphi)=-\frac{e}{\hbar}\sum_{n,E_n>0}\tanh(\frac{E_n}{2k_BT})\pdv{E_n}{\varphi} , \label{Eq:CurrentFreeEnergy}
\end{align}
where $E_n$ are the energies of the Hamiltonian as defined in Eq.~\eqref{eq:FullTightBindingHamiltonian} \cite{Cayao2017Majorana}. 

We note that the code performance is better in case of the first method based on the Green's function calculated in the infinite system compared to the second method,  which is based on the eigenvalue calculation in finite-size systems. Therefore, most of the current calculations are based on the first method. 

Finally, we introduce the directional dependent critical currents $I_{+}^c$ and $I_{-}^c$ that are the maxima and minima of the current phase relation for all phases $\varphi \in [0, 2\pi)$. These represent the critical current for current flow to the right and left, respectively. The corresponding diode efficiency is defined as 
\begin{align}
\eta=\frac{I_{+}^c-|I_{-}^c|}{(I_{+}^c+|I_{-}^c|)/2}.
\end{align}

\section{SDE for uniform ferromagnetic exchange coupling\label{Sec:SDEForFerromagneticSystems}}
\subsection{Dependence on the chemical potential}
\begin{figure*}[t]
\subfloat{\label{figFerroMagChemPotCurrent}\stackinset{l}{-0.00in}{t}{-0.0in}{(a)}{\stackinset{l}{1.8in}{t}{0in}{(b)}{\stackinset{l}{3.6in}{t}{-0.0in}{(c)}{\stackinset{l}{5.4in}{t}{0.in}{(d)}{\stackinset{l}{-0.00in}{t}{2.in}{(e)}{\stackinset{l}{1.8in}{t}{2.in}{(f)}{\stackinset{l}{3.6in}{t}{2.in}{(g)}{\stackinset{l}{5.4in}{t}{2.in}{(h)}{\includegraphics[width=1\textwidth]{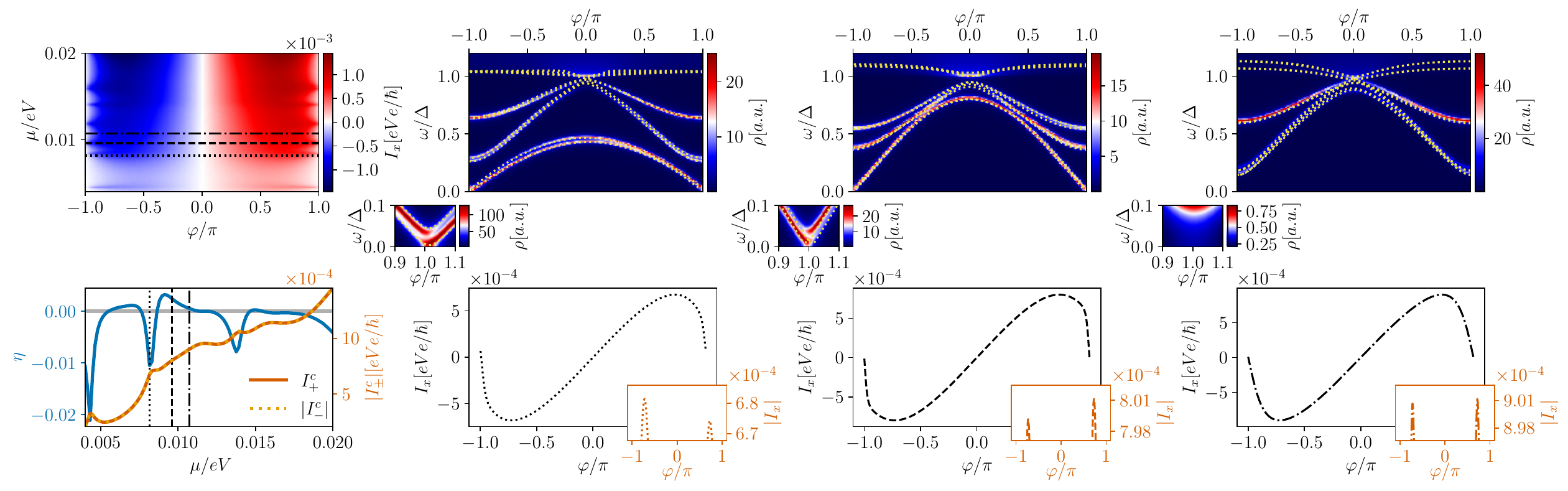}}}}}}}}}
}
\subfloat{\label{figLDOS_NegDiodeEff}}
\subfloat{\label{figLDOS_PosDiodeEff}}
\subfloat{\label{figLDOS_ZeroDiodeEff}}
\subfloat{\label{figDiodeEffAsFuncOfMu}}
\subfloat{\label{figCPR_NegDiodeEff}}
\subfloat{\label{figCPR_PosDiodeEff}}
\subfloat{\label{igCPR_ZeroDiodeEff}}
\caption{ \textit{SDE as a function of the chemical potential in case of an in-plane ferromagnetic texture}: (a) Supercurrent $I_x$ as a function of the chemical potential $\mu$ and of the superconducting phase difference $\varphi$. (b-d)  The LDOS,  $\rho$, as a function of the superconducting phase difference $\varphi$. The values chosen for the chemical potentials are indicated by the (b) dotted, (c)  dashed, and (d)  dashed-dotted  lines in panel (a). We compare the LDOS calculated at a single site at the end of the junction  (with the position of the site chosen similar  to the green encircled site in Fig.~\ref{figSchematicsLattice2}) found via Green's function method in the infinite system (with leads) with the low-energy energy spectrum (yellow dashed lines) obtained for a finite-size system in which the superconducting regions are longer than the coherence length (no superconducting leads).  If the lowest energy state crosses zero energy, then the diode efficiency is non-zero, see panel (b) and (c). In contrast, if the lowest state does not cross zero energy, then the diode efficiency approaches almost zero, see panel (d). (e) Left [right] $y$-axis: diode efficiency [critical currents] as a function of the chemical potential. Here, the diode efficiency changes its sign, meaning that the diode can switch its polarity as a function of the chemical potential. (f-h) CPR for the position of the chemical potentials shown as in panels (b-d). The inset shows the modulus of the current  and it reveals the diode polarity, which can hardly be read off from the bare CPR. Parameters: $L_y=2L_x=140$ nm, $m_{\text{eff}}=0.023m_e$, $a=2.5$ nm, $\Delta=0.8$ meV, $J=0.05$ meV, $\alpha=0.05$ eVnm, and $\gamma=32$~meV.}
 \label{figParameterOptimization1}
\end{figure*}

In this section, we study the diode efficiency for a system with a uniform ferromagnetic exchange coupling oriented in parallel to the junction ($y$-direction in Fig.~\ref{figSchematicsLattice2}). This is done in order to find an optimal parameter range for the operation of the SC diode. 

We note that, in principle, all ingredients for Majorana bound states (MBSs) in a planar Josephson junction are present, namely (Rashba) SOI due to broken inversion symmetry and exchange coupling, which acts as a local magnetic field, and a superconducting phase difference. Consequently, the appearance of a topological phase is a question of the chosen parameters  \cite{Pientka2017Topological, Hell2017TwoDimensional,Fornieri2019Evidence, Ren2019Topological, Luethi2023Planar}.  In this work, however, we focus on the case of $L_y\ll \xi_{sc}$, where $\xi_{sc}$ denotes the superconducting coherence length, so that the system does not host well localized MBSs.
We start our analysis by the calculation of the current as a function of the superconducting phase and the chemical potential, see Fig.~\ref{figFerroMagChemPotCurrent}. We find that the current reveals oscillations as a function of the chemical potential. Additionally, the critical currents increase with growing $\mu$, which can be partially explained with a higher transparency at larger $\mu$. We estimate for the chosen parameters an average transparency of $\tau=0.77$  and $\tau=0.94$ at $\mu/\Delta=10$  in absence of the exchange coupling for the cases $E_{so}/\Delta=0$ 
and  $E_{so}/\Delta=0.47$, respectively. In particular, we fitted the current phase relation (CPR) with the formula \cite{Titov2006Josephson}
\begin{align}
I_x(\varphi)=\frac{A\sin(\varphi)}{\sqrt{1-\tau\sin ^2\left(\varphi/2\right)}},
\end{align}
where $A$ and $\tau$ serve as fit parameters \footnote{The additional factor of  $\tau$ in Ref. \cite{Titov2006Josephson},  has been absorbed into the amplitude $A$.}.

We choose three values of the chemical potential, see the dotted, dashed and dotted-dashed lines in Fig.~\ref{figFerroMagChemPotCurrent}, and plot the corresponding current phase relation in Figs.~\ref{figCPR_NegDiodeEff}-\ref{igCPR_ZeroDiodeEff}. The inset shows the modulus of the current, highlighting that there is a difference of the critical currents and also that this difference in critical currents depends on chemical potential. Here, the diode efficiency $\eta$ is quite small due to the weak exchange coupling. This small exchange coupling was chosen in order to reduce the phase space of the topological phase. In general, if the transparency of the junction is reduced, then the topological phase shrinks for fixed finite exchange couplings \cite{Pientka2017Topological}. However, as mentioned above, well localized Majorana bound states cannot form in junctions where the length ($y$-direction of Fig.~\ref{figSchematicsLattice2}) is short.

We also calculated the LDOS at one site located at the end of the junction (see e.g. the green encircled site in Fig.~\ref{figSchematicsLattice2}) by attaching superconducting leads as described in the Sec.~\ref{App:CurrCalc}. Here, we show only the positive energy range, see Figs.~\ref{figLDOS_NegDiodeEff}-\ref{figLDOS_ZeroDiodeEff}, a comparison with  the energy spectrum (yellow dashed lines) calculated in a finite size system, in which we discretized the superconducting regions (no superconducting leads), reveals a good agreement.  We find that the diode efficiency decreases when the lowest ABS is pushed to higher energies for superconducting phases close to $\pi$. A magnification of the low energy region reveals in particular that the lowest ABS energy is almost a linear function of $\varphi$ close to $\varphi=\pi$ for systems with sizeable diode efficiency, indicating a high transparency mode \cite{Baumgartner2022Effect}. In contrast, if the diode efficiency is almost zero, then the derivative of the lowest ABS energy vanishes close to  $\varphi=\pi$. Finally, we calculated the diode efficiency and the critical currents as a function of the chemical potential, see Fig.~\ref{figDiodeEffAsFuncOfMu}. Notably, the diode efficiency changes its sign multiple times, we attribute this behaviour partially to the behaviour of the lowest ABS, which is strongly influenced by the choice of the chemical potential. In principle, such a gate tunable SDE can be used as a Josephson transistor \cite{mazur2022gatetunable}, below we will show that this is also possible simply by moving a magnetic texture along the racetrack.\\

\begin{figure*}[t]
\subfloat{\label{figFerroMagChemPotCurrent_ExCoup}\stackinset{l}{-0.00in}{t}{-0.0in}{(a)}{\stackinset{l}{1.8in}{t}{0in}{(b)}{\stackinset{l}{3.6in}{t}{-0.0in}{(c)}{\stackinset{l}{5.4in}{t}{0.in}{(d)}{\stackinset{l}{-0.00in}{t}{2.in}{(e)}{\stackinset{l}{1.8in}{t}{2.in}{(f)}{\stackinset{l}{3.6in}{t}{2.in}{(g)}{\stackinset{l}{5.4in}{t}{2.in}{(h)}{\includegraphics[width=1\textwidth]{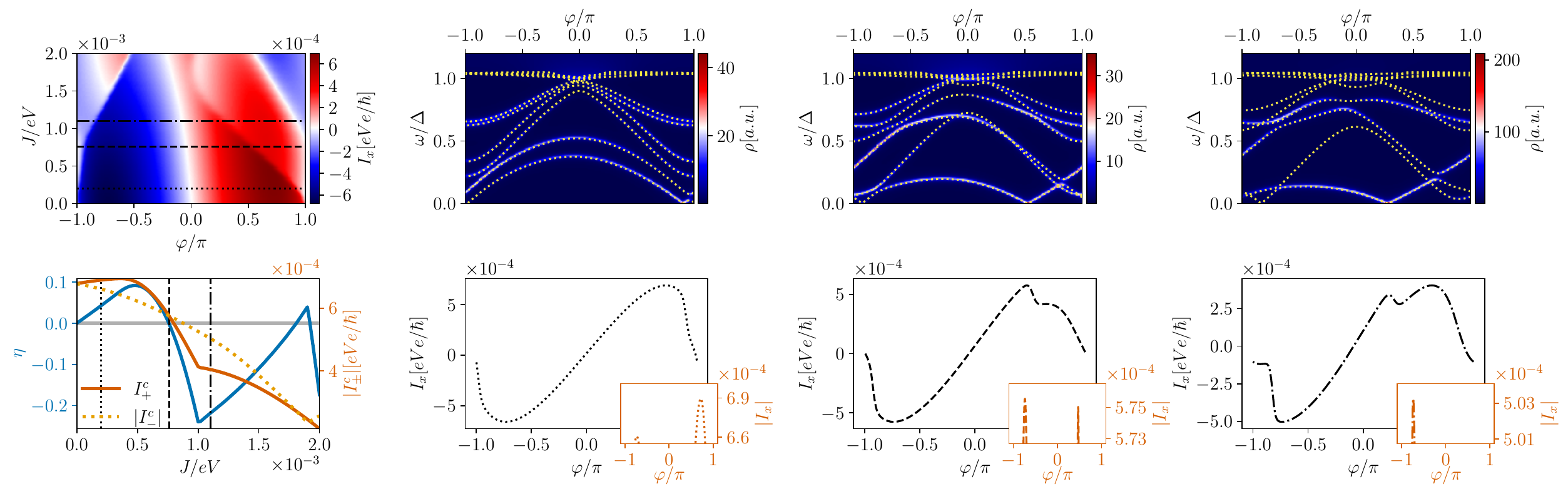}}}}}}}}}
}
\subfloat{\label{figLDOS_NegDiodeEff_ExCoup}}
\subfloat{\label{figLDOS_PosDiodeEff_ExCoup}}
\subfloat{\label{figLDOS_ZeroDiodeEff_ExCoup}}
\subfloat{\label{figDiodeEffAsFuncOfMu_ExCoup}}
\subfloat{\label{figCPR_NegDiodeEff_ExCoup}}
\subfloat{\label{figCPR_PosDiodeEff_ExCoup}}
\subfloat{\label{figCPR_ZeroDiodeEff_ExCoup}}
\caption{ \textit{SDE as a function of the exchange coupling strength  in case of an in-plane ferromagnetic texture}: (a) Supercurrent $I_x$ as a function of the exchange coupling $J$ and of the superconducting phase difference $\varphi$. (b-d)  The LDOS,  $\rho$, as a function of the superconducting phase difference $\varphi$. The values chosen for the exchange coupling are indicated by the (b) dotted, (c)  dashed, and (d) dashed-dotted  lines in panel (a). We compare the LDOS at a single site at the end of the junction  (with the position of the site chosen similar  to the green encircled site in Fig.~\ref{figSchematicsLattice2}), calculated by using Green's function method in the infinite system (with leads) with the low energy spectrum (yellow dashed lines) of a finite-size system in which the superconducting regions are longer than the coherence length (no superconducting leads). The zero-energy crossing of the lowest energy state can lead to kinks in the CPR.  (d) Left [right] $y$-axis: diode efficiency [critical currents] as a function of the exchange coupling. The diode can change its polarity. (f-h) The CPR for the exchange couplings as in panels (b-d). Parameters: $L_y=2L_x=140$ nm, $m_{\text{eff}}=0.023m_e$, $a=2.5$ nm, $\Delta=0.8$ meV, $\alpha=0.05$ eVnm, $\mu=8.16$ meV, and $\gamma=32$~meV.}
 \label{figParameterOptimization_ExCoup1}
\end{figure*}

\subsection{Dependence on the exchange coupling $J$}
Next, we study the current as a function of the exchange coupling strength $J$ for the same ferromagnetic texture, i.e., pointing in $y$-direction, see Fig.~\ref{figParameterOptimization_ExCoup1}. The critical currents decrease with growing exchange coupling strength. 
However, in general, the overall behaviour of the current is quite complicated due to the low energy sub-gap states, which we analyze in Figs.~\ref{figLDOS_NegDiodeEff_ExCoup}-\ref{figLDOS_ZeroDiodeEff_ExCoup} for three different values of $J$ as indicated by the dotted, dashed and dashed-dotted line in panel Fig.~\ref{figFerroMagChemPotCurrent_ExCoup}. If an ABS crosses zero energy, then the (central) derivative of the ABS energy with respect to the superconducting phase difference is not well defined since only the negative eigenvalues contribute to the ground state and therefore to the current phase relation. A different sign of left and right derivative can lead to jumps in the CPRs, see Eq.~\eqref{Eq:CurrentFreeEnergy} and Figs.~\ref{figCPR_NegDiodeEff_ExCoup}-\ref{figCPR_ZeroDiodeEff_ExCoup}. These jumps, in turn, lead to strong changes in the diode efficiency including sign changes, as can be read out from the insets, which show the absolute value of the currents. The overall behaviour of the diode efficiency  as a function of the exchange coupling is shown in Fig.~\ref{figDiodeEffAsFuncOfMu_ExCoup}. For small values of $J $, the diode efficiency $\eta$ increases approximately linearly with exchange coupling strength. In contrast, for large exchange coupling strengths, the diode efficiency deviates from the linear behaviour and can even change its sign. In the linear regime, the lowest state crosses zero energy close to $\varphi=\pi$. At the exchange coupling associated with the sign change of the diode efficiency, the zero-energy crossing of the lowest state is pushed away from $\varphi=\pi$. Finally, we note that the SDE generally increases substantially for larger exchange couplings, in part due to the smaller critical currents, as long as the system is not fine-tuned to a chemical potential where the SDE vanishes completely.

\subsection{Local supercurrents}
\begin{figure}[t]
\subfloat{\label{figFermiOsciOfLocalCurrent}\stackinset{l}{-0.00in}{t}{-0.0in}{(a)}{\stackinset{l}{1.7in}{t}{-0.05in}{(b)}{\includegraphics[width=1\columnwidth]{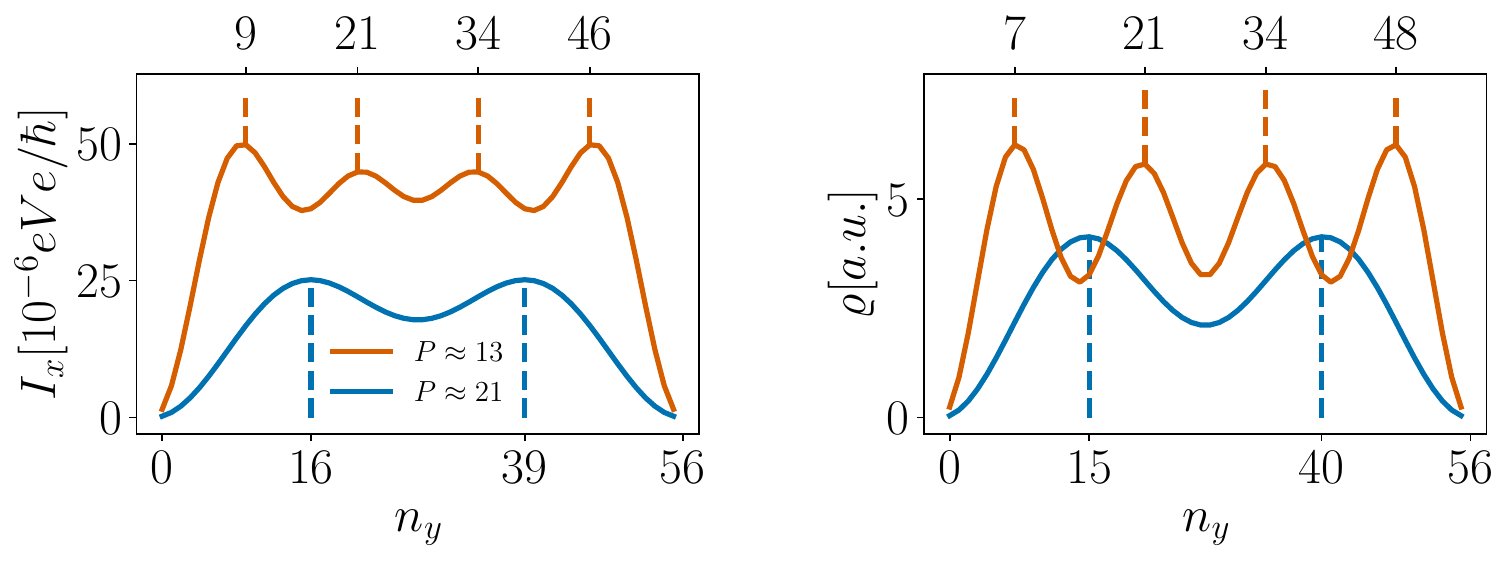}}}
}
\subfloat{\label{figFermiOsciOfLDOS}}
\caption{ 
\textit{Spatial oscillations in the distribution of local supercurrents:} (a) Local currents as a function of the $y$-coordinate $n_y$. The red and blue curve correspond to systems with different chemical potential and therefore different Fermi momenta, which determine the oscillation period $P$ (in units of $a$).
 We find that the spatial separation of the peaks agrees well with the analytic prediction of the oscillation period set by the Fermi wavelength. 
 (b)   Energy-integrated LDOS [see Eq.~\eqref{Eq:IntegratedLDOS}] along a column of sites to the left of the cross section through which we calculated the current. The oscillation profile of LDOS matches quite well with the profile of the local supercurrents.  
 Parameters: $L_y=2L_x=140$ nm, $m_{\text{eff}}=0.023m_e$, $a=2.5$ nm, $\Delta=0.8$ meV, $J=0$ meV, $\alpha=0$ eVnm, $\gamma=0$ meV,  and $\mu=15$ ($\mu=6$) meV for the red (blue) graph. }
 \label{figCurrentAndDiodeEffAsFuncOfChemPot}
\end{figure}

So far, we have only considered the total supercurrent flowing in $x$-direction through the junction. Here, in contrast, we analyze the local supercurrents in $x$-direction as a function of the $y$-coordinate. With respect to Fig.~\ref{figSchematicsLattice2}, this means that we study the current on individual red bonds. In order to simplify the analysis, we set the  exchange coupling $J$ and the Rashba SOI $\alpha$ to zero, such that there is no SDE. It turns out that the current strength oscillates along the $y$-direction with an approximate period $\lambda_F /2$ set by the Fermi wavelength $\lambda_F\equiv 2\pi/k_F$, where $k_F$ is the Fermi momentum. 
In Fig.~\ref{figFermiOsciOfLocalCurrent} we analyze these current oscillation for two different values of the chemical potential. 

To connect the LDOS to the current, we note that Eq.~\eqref{Eq:CurrentfromGreensFunc} can be rewritten in terms of the eigenenergies and wavefunctions. This explains the oscillatory behavior of the current with respect to chemical potential, since the wave functions of the ABSs exhibit oscillations in $y$-direction set by the Fermi wavelength. In order to quantify the direct correlation between oscillations of the wave functions and of the current, we integrate the LDOS over energy window inside the superconducting gap:
\begin{align}
\varrho(\mathbf{n})=\int_0^{\Delta} \rho(\omega,\mathbf{n}) \dd \omega, \label{Eq:IntegratedLDOS}
\end{align}
 which also captures the dependence on the superconducting phase difference. In terms of the schematic picture shown in  Fig.~\ref{figSchematicsLattice2}, this means that we consider the LDOS along the left column of blue sites. The integrated LDOS, which takes all sub gap states into account, reveals a similar oscillation pattern as in the current, see Fig.~\ref{figFermiOsciOfLDOS}.

Finally, we note that the oscillations of the current as a function of position can affect the critical currents in systems with finite-size magnetic texture like domain walls or skyrmions. In fact, the ratio between the Fermi wavelength and the spatial extent of the magnetic defect, in our case $\lambda_s a$ or $\lambda_{dw} a$, plays a central role. For example, if $\lambda_s a \gg \lambda_F $, then the effect of the spatial oscillation gets averaged out and have less impact on the diode efficiency as a function of position of the magnetic texture.

\section{SDE for a texture moving on a racetrack \label{Sec:ScDiodeEffectOnARaceTrack}}
We now consider what happens to the SDE when a given magnetic texture moves along the portion of the racetrack that forms the normal section of the Josephson junction. We will see that the nature of the magnetic texture and its position within the junction can significantly modify the diode efficiency, $\eta$, and even change its sign. As a result, the magnetic texture can be detected by these modifications in the SDE as it moves through the junction or, conversely, moving a magnetic texture through the junction can be used to change the sign of the diode efficiency and therefore create a Josephson transistor effect.

\subsection{Domain walls}
\begin{figure}[t]
\subfloat{\label{figCurrentMag1}\stackinset{l}{-0.in}{t}{-0.0in}{(a)}{\stackinset{l}{1.7in}{t}{-0.05in}{(b)}{\stackinset{l}{-0.00in}{t}{0.9in}{(c)}{\stackinset{l}{1.7in}{t}{0.9in}{(d)}{\includegraphics[width=1\columnwidth]{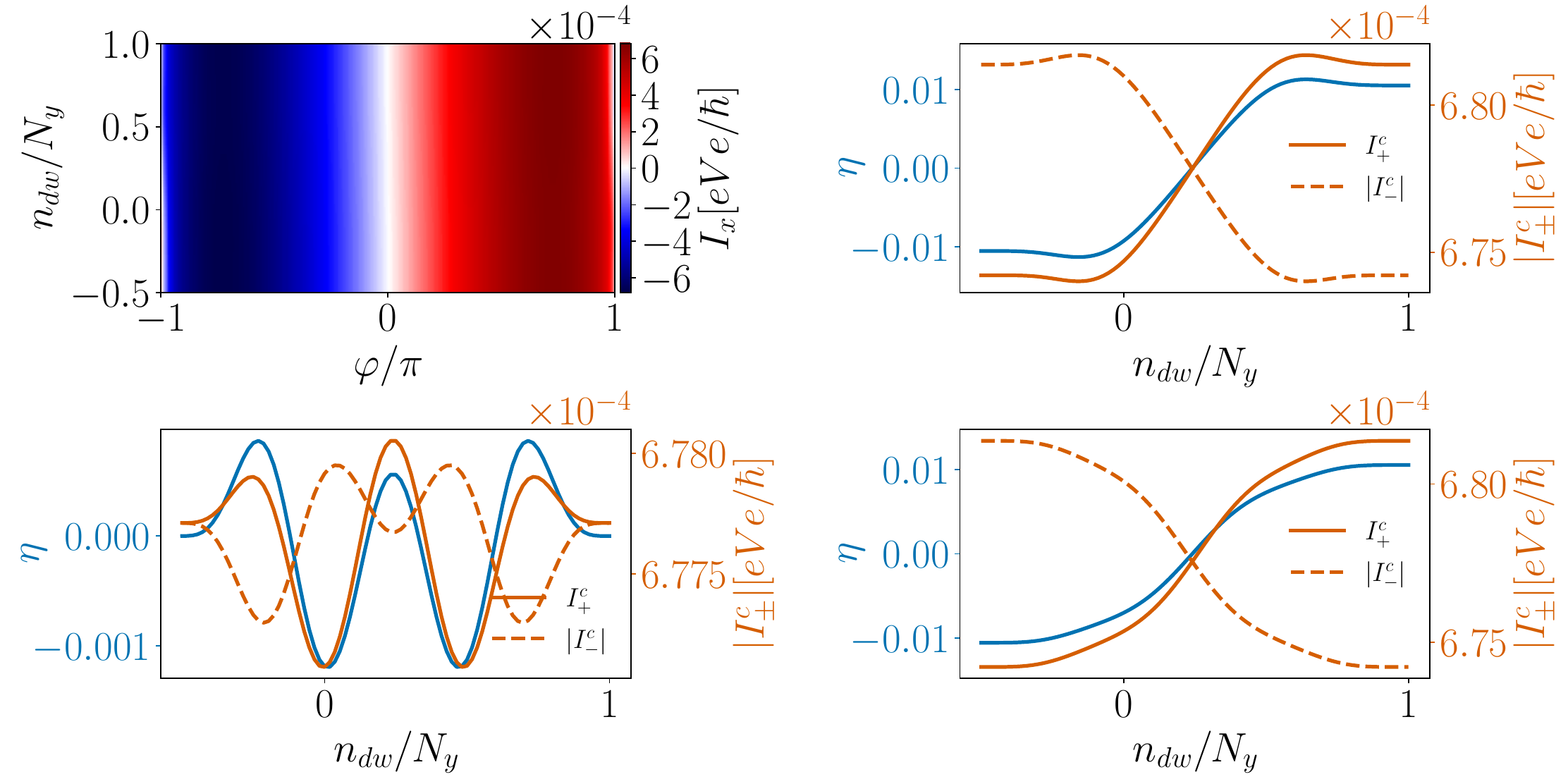}}}}}
}
\subfloat{\label{figDiodeEffMag1}}
\subfloat{\label{figDiodeEffMag2}}
\subfloat{\label{figDiodeEffMag3}}
\caption{ \textit{The CPR and diode efficiency as a function of the position of the domain wall:} (a) Supercurrent as a function of phase difference $\varphi$ and the position of the domain wall $n_{dw}$. Panels (a) and (b) correspond to the domain wall profile $\Phi_{dw,1}$, while panels (c) and (d), correspond to the profiles $\Phi_{dw,2}$ and $\Phi_{dw,3}$, respectively.  (b,c,d) Critical currents $I^c_{\pm}$ and diode efficiency $\eta$ as a function of  $n_{dw}$. The diode changes its sign for the magnetic textures considered in (a,b) and (d). In contrast, when the magnetization is out of plane far away from the domain wall (c), then the SDE appears only when the domain wall enters the junction and we observe additional oscillations of the diode efficiency when the domain wall moves through the system.   Parameters: $L_y=2L_x=140$ nm, $m_{\text{eff}}=0.023m_e$, $\mu=8.16$ meV, $a=2.5$ nm, $\Delta=0.8$ meV, $J=0.05$ meV, $\alpha=0.05 $ eVnm,  $\gamma=32$ meV, and $\lambda_{dw} a=70$ nm.   }
 \label{figDomainWallsAndDiode}
\end{figure}

We first study the SDE due to a magnetic domain wall moving through the Josephson junction. First, we consider a magnetic texture as defined by the angle profiles $\Phi_{dw,1}$ and $\Theta_{dw,1}$ and calculate the current as a function of the superconducting phase difference and of the position of the domain wall, see Fig.~\ref{figCurrentMag1}. The phases associated with the positive and negative critical currents as well as the phase associated with zero current change as a function of the position of the domain wall. More importantly, the direction of the exchange field of the magnetic texture reverses when the domain wall passes through the junction and, consequently, the direction of the SDE also inverts, resulting in a Josephson transistor effect. This behaviour manifests itself in a diode efficiency that changes its sign when the domain wall passes through the center of the system ($\approx N_y/2$), see Fig.~\ref{figDiodeEffMag1}. The sign change and the value $\eta=0$ for a system with the domain wall in the middle is enforced by symmetry. Moreover, we note that as the domain wall moves through the system, the magnitude of the critical currents change significantly.

In general, the physics of the system are strongly determined by the magnetization direction: Repeating the same calculation with an out-of-plane magnetization far away from the domain wall, as defined by $\Phi_{dw,2}$ and $\Theta_{dw,2}$, reveals a different behavior, see Fig.~\ref{figDiodeEffMag2}. In particular, if the domain wall is far away from the junction, then the out-of-plane magnetization does not result in a diode effect and consequently the diode efficiency is zero.

The overall behaviour of the diode efficiency as a function of the domain wall position exhibits several sign changes. In a simple picture, one might expect that the diode efficiency first increases when the domain wall moves into the junction until the whole domain wall entered the system.
The efficiency would be constant until the domain wall starts to leave the junction. However, the calculated diode effect exhibits a more complex behaviour, see Fig.~\ref{figDiodeEffMag2}. The diode efficiency does at first grow and is almost constant when the domain wall is located in the middle of the junction, due to the symmetry of the system. However, $\eta$ exhibits in total four sign changes as the domain wall moves through the junction. We attribute this behaviour to the wavefunctions of the ABSs in the junction. As discussed above, the ABSs in the junction exhibit spatially dependent oscillations along the $y$-direction, resulting in both a probability density and current changes that depends locally on the $y$-coordinate. Consequently, the diode efficiency can change, including sign changes and associated Josephson transistor effect, if the spatial extent of the domain wall $\lambda_{dw} a$ is of the same order as the Fermi wavelength $\lambda_F$. Although for numerical ease we utilized a low chemical potential with large $\lambda_F$, in practice, many proposed racetrack materials are good metals \cite{Ryu2013chiral, Parkin2015Memory} such that $\lambda_F\gg \lambda_{dw} a$ and these multiple sign changes would not be expected.

Next, we note that the magnitude of the diode efficiency, $\eta$, for the particular choice of magnetization set by $\Phi_{dw,2}$ and $\Theta_{dw,2}$, strongly depends on the ratio between the length of the domain wall and the length of the junction, which are set by $\lambda_{dw}$ and  $N_y$, respectively. In particular, the longer the junction compared to the domain wall the smaller the superconducting diode efficiency of the whole junction. This is why we chose junctions which are just a few times longer than $\lambda_{dw}$ and avoid the regime $N_y\gg \lambda_{dw}$.  

Finally, we studied a junction with a magnetization as defined via  $\Phi_{dw,3}$ and $\Theta_{dw,3}$, see Fig.~\ref{figDiodeEffMag3}. This junction behaves similar to the first considered set-up with $\Phi_{dw,1}$ and $\Theta_{dw,1}$. In particular, $|\eta|$ 
is constant until the domain wall enters the junction, then it decreases until the domain wall reaches the center of the junction at which point a sign change in $\eta$ occurs and $|\eta|$  increases until the domain wall exits the junction. 

\subsection{Skyrmions}
\begin{figure}[t]
\subfloat{\label{figDiodeEffSkyr1}\stackinset{l}{-0.00in}{t}{-0.0in}{(a)}{\stackinset{l}{1.7in}{t}{-0.05in}{(b)}{\includegraphics[width=1\columnwidth]{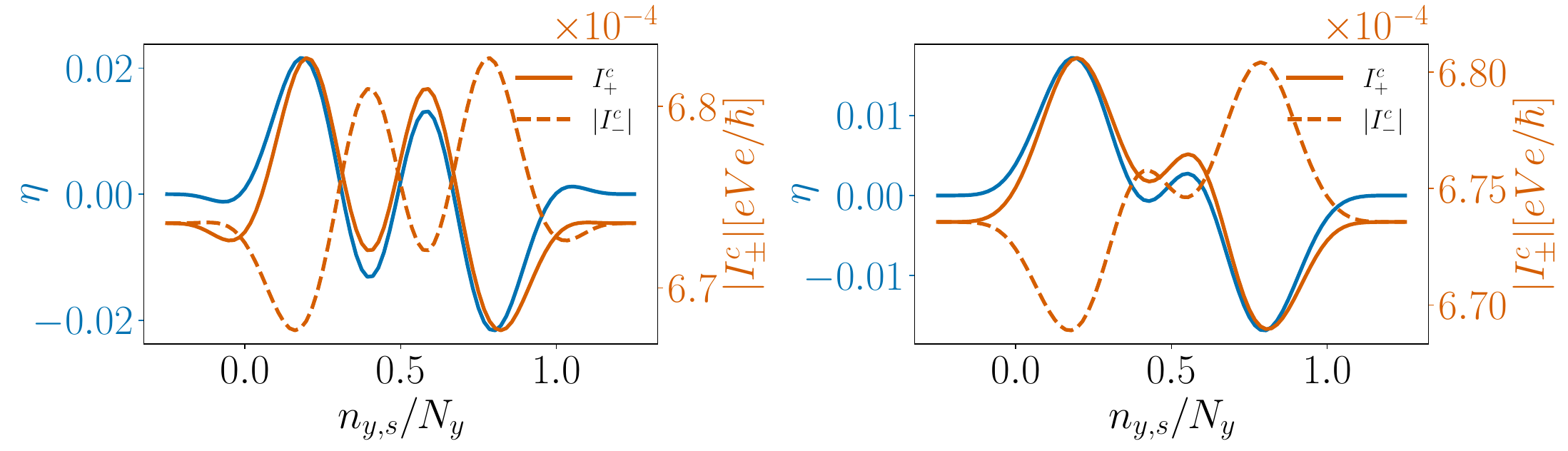}}}
}
\subfloat{\label{figDiodeEffSkyr2}}
\caption{ \textit{ Diode efficiency as a function of the position of the skyrmion:} Critical currents $I^c_{\pm}$ and diode efficiency $\eta$ as a function of  $n_{y,s}$ for (a) [(b)]:  Néel [Bloch] skyrmion.  The diode efficiency strongly depends on the position of the Néel or Bloch skyrmion and there are certain positions, depending on the system configurations, where $|\eta|$ is maximized.  If there is no skyrmion inside the junction, then there is no SDE ($\eta=0$) since the background magnetization is out of plane. 
 Parameters: $L_y=2L_x=140$ nm, $m_{\text{eff}}=0.023m_e$, $\mu=8.16$ meV, $a=2.5$ nm, $\Delta=0.8$ meV, $J=-0.2$ meV, $\alpha=0.05 $ eVnm, and $\gamma=32$ meV, and $\lambda_{s}a=35$ nm.}
 \label{figSkyrmionsAndDiode}
\end{figure}

We now repeat a similar analysis for the domain wall setups studied above but instead for racetracks hosting skyrmions. First, we note that we only consider skyrmions with a ferromagnetic background aligned in $z$-direction (out of plane). Therefore, there is no diode effect if a skyrmion is not in the junction ($n_{y,s}\leq - \lambda_s/2$), see Fig.~\ref{figSkyrmionsAndDiode}. Considering first Néel skyrmions moving on a racetrack, when the skyrmion enters the junction the tilted magnetization close to the skyrmion core leads to a finite SDE, see Fig.~\ref{figDiodeEffSkyr1}. The strength of this effect strongly depends on the ratio between $\lambda_s$ and $N_y$, as in the case of the second domain wall configurations with $\Phi_{dw,2}$ and $\Theta_{dw,2}$ studied above. As above, these spatial oscillations of $\eta$ are set by the Fermi wavelength and result in a complicated behaviour that can exhibit several sign changes of $\eta$, see Fig.~\ref{figDiodeEffSkyr1}.

Finally, we consider a Bloch skyrmion with a magnetization as defined in Sec.~\ref{Sec_SubSec_Skyrmions}, see Fig.~\ref{figDiodeEffSkyr2}.  The general behaviour is quite similar to the Néel skyrmion and in general we do not find a substantial difference in the diode efficiency response caused by the two types of skyrmionic texture.

\section{SDE in Josephson junctions without Rashba SOI \label{Sec_SDE_Without_ExplicitSOI}} 
So far we have explicitly incorporated Rashba SOI in our model via Eq.~\eqref{eq:HamSOI}. Here, instead, we remove the Rashba SOI and investigate which type of magnetic texture can support a SDE by itself. In order to answer this question, we first consider the  continuum Hamiltonian 
\begin{equation}
H=\int \dd x\ \dd y \ \Psi^{\dagger} (\vec r) \mathcal{H}(\vec r)\Psi (\vec r),
\end{equation} where $\Psi^{\dagger}(\vec r)=(\Psi_{ \vec r,\uparrow}^{\dagger},\Psi_{ \vec r,\downarrow}^{\dagger},\Psi_{ \vec r,\downarrow}, -\Psi_{\vec r,\uparrow})$ is a vector composed of  the field operators $\Psi_{\vec r,\uparrow}^{\dagger}$ ($\Psi_{\vec r,\uparrow}$) and $\Psi_{\vec r,\downarrow}^{\dagger}$  ($\Psi_{\vec r,\downarrow}$) which create (annihilate) a particle at the position $\vec r=(x,y)$ with spin up or down, respectively. The Hamiltonian density is given by 
\begin{align}
&\mathcal{H}(\vec r;\varphi_{\vec r})= - \frac{\hbar^2}{2m}(\nabla_x^2+\nabla_y^2)\tau_z \label{Eq:MagneticTextureContinuumModel}\\ 
&+\Delta_{\vec r}( \tau_x\cos \varphi_{\vec r} +\tau_y\sin\varphi_{\vec r})
+J \tau_0 \vec S_{\vec r}(\Phi_{\vec r},\vartheta_{\vec r}) \cdot \vec \sigma \nonumber,
\end{align}
where $\sigma_j$ and $\tau_j$ are Pauli-matrices acting in spin space and particle hole space respectively, and $\Delta_{\vec r}$ is real. As in Eq.~\eqref{eq:MagnetizationVector}, the spin texture is finite only within the normal section and using spherical coordinates such that $\vec S_{\vec r}(\Phi_{\vec r},\vartheta_{\vec r})=[\cos(\vartheta_{\vec r})\sin(\Phi_{\vec r}),\sin(\vartheta_{\vec r})\sin(\Phi_{\vec r}),\cos(\Phi_{\vec r})]$ and $\vec \sigma=(\sigma_x,\sigma_y,\sigma_z)$ the vector of Pauli matrices. Both the angles $\Phi_{\vec r}$ and $\vartheta_{\vec r}$ as well as the superconducting phase difference $\varphi_{\vec r}$ can depend on position $\vec r=(x,y)$ of a given spin of the texture.

Although we are interested in cases where an SDE does occur in the absence of Rashba SOI, we first point out that there are several spin textures where symmetries still do not allow the system to support an SDE. For instance, if the angles $\vartheta_{\vec r}$ and $\Phi_{\vec r}$ are constant then all spins are parallel, so the system breaks only time reversal symmetry. The absence of Rashba SOI in the 2DEG means that there is no coupling between spin-space and real space, allowing for arbitrary rotations in spin space. As a consequence, the combination of time-reversal symmetry, $\mathcal{T}=i \sigma_y \tau_0 \mathcal{K}$, where $\mathcal{K}$ is complex conjugation, with a rotation in spin-space by $\pi$ around an axis in the plane perpendicular to the direction of the spins, $\mathcal{R}_\pi$, results in the identity $\mathcal{R}_\pi\mathcal{T}\mathcal{H}(\vec r;\varphi_\vec r)(\mathcal{R}_\pi \mathcal{T})^\dagger=\mathcal{H}(\vec r;-\varphi_\vec r)$. This implies that all eigenenergies satisfy $E_n(\varphi)=E_n(-\varphi)$, however, since the current is given by the derivative of eigenenergies with respect to the phase difference, see Eq.~\eqref{Eq:CurrentFreeEnergy}, this identity means that the current must satisfy $I(\varphi)=-I(-\varphi)$ and therefore no SDE can occur since $I_+^c=|I_-^c|$.

In fact, more generally, if all spins lie in a plane, then the combination of a $\pi$ rotation around the axis defining the plane and time reversal symmetry will result in the same identity and the absence of an SDE. For instance, if all spins lie in the $xz$-plane then a rotation in spin-space about the $y$-axis, such that $\mathcal{R}_\pi=\sigma_y$, will also result in the identity $\mathcal{R}_\pi\mathcal{T}\mathcal{H}(\vec r;\varphi_\vec r)(\mathcal{R}_\pi \mathcal{T})^\dagger=\mathcal{H}(\vec r;-\varphi_\vec r)$ which ensures the absence of an SDE. As such, in the absence of Rashba SOI, a simple domain wall will not result in an SDE since it only rotates in a single plane.

Additionally, we note that spatial symmetries can also result in a similar identity that will forbid an SDE. Namely, for our setup we are interested in currents in the $x$-direction i.e. across the junction. If the spin-texture, $\vec S_{\vec r}(\Phi_{\vec r},\vartheta_{\vec r})$, only depends on the $y$-coordinate then the transformation $\vec r = (x,y) \rightarrow \vec r'=(-x,y)$ gives again $H(\vec r';\varphi_{\vec r'})=H(\vec r ;-\varphi_{\vec r})$, where we used the fact the phase $\varphi_{\vec r}$ only varies in the $x$-direction and that, without loss of generality, the phase can be taken to be of opposite in sign in the left and right superconducting sections of the junction.

We now demonstrate that an SDE is allowed in the absence of Rashba SOI. First, since the superconducting terms are unaffected by the following transformations and for the ease of discussion, we set $\Delta_{\vec r}=0$ everywhere in the BdG Hamiltonian presented in Eq.~\eqref{Eq:MagneticTextureContinuumModel}. Next, apply the gauge transformation $U_1=e^{-i(\vartheta_{\vec r}/2-\pi/4)\sigma_z}$ via $\tilde{\mathcal{H}}=U_1^{\dagger}\mathcal{H} U_1$. This transformation rotates the magnetic texture around the $z$-axis into the $yz$-plane. Subsequently, we apply the second gauge transformation $U_2=e^{i(\Phi/2-\pi/4)\sigma_x}$  as $\mathcal{H}'=U_2^{\dagger}\tilde{\mathcal{H}}U_2$ and obtain
\begin{align}
\mathcal{H}'=&\frac{-\hbar^2}{2m}\sum_{x_j \in{\{x,y\}}}\Bigg[ \bigg(\pdv[2]{x_j}-\frac{1}{4}\left\lbrace\pdv{\vartheta}{x_j}\right\rbrace^2-\frac{1}{4}\left\lbrace\pdv{\Phi}{x_j}\right\rbrace^2\bigg)\sigma_0  \nonumber \\
-i &\left\lbrace\pdv{x_j}\Lambda_{x_j}(x,y)+\Lambda_{x_j}(x,y)\pdv{x_j}\right\rbrace
\Bigg]+J \sigma_y, \label{eq:TransformedHamEffSOIAndUniformField}
\end{align}
where
\begin{align}
\Lambda_{x_j}(x,y)=\frac{1}{2}\left[\cos(\Phi)\sigma_y+\sin(\Phi)\sigma_z\right]\pdv{\vartheta}{x_j}-\frac{1}{2}\pdv{\Phi}{x_j}\sigma_x,
\end{align}
see Appendix~\ref{Sec:AppGaugeTrafo} for a detailed derivation. In this new basis, after the unitary transformation has been applied, the exchange coupling term, namely the term proportional to $J$, is ferromagnetic. This modification of the exchange coupling field, however, results in new effective SOI term and a non-uniform chemical potential \cite{Braunecker2010SpinSelective, Choy2011Majorana, Martin2012Majorana, Perge2013Proposal, Nakosai2013TwodDimensional, Chen2015Majorana, Hess2022Prevalence}. As discussed extensively above, SOI and a ferromagnetic background coupling to the direction of that SOI is the key ingredient that can enable an SDE. Given that a sufficiently complex spin-texture transforms to an effective SOI with ferromagnetic background, it is clear that this can result in an SDE, as long as it is not symmetry forbidden. In Table~\ref{Tab:SDE_DueToMagTextures} we give all possible forms of spin-textures and indicate whether or not an SDE is allowed in the absence of Rashba SOI as well as whether or not the symmetries discussed above are present. We note, even if allowed by the general symmetries discussed above, fine-tuning of the position of magnetic textures can result in the absence of an SDE due to, e.g., spatial rotational symmetries.

\begin{table}[tb]
 \centering
\begin{tabular}{c|c||ccc}
\hline
\hline
$\Phi$ &$\vartheta$ & $\mathcal{R}_\pi\mathcal{T}$ & $x\rightarrow-x$ & $\mathrm{SDE}[I_x]$ \\
\hline
$\Phi=\frac{\pi}{2}$ & $\vartheta =\rm{const.}$ & $\checkmark$ & $\checkmark$ & \color{red}$\times$\color{black}  \\ 
$\Phi=\frac{\pi}{2}$ & $\vartheta =\vartheta(x)$ & $\checkmark$ & $\times$ & \color{red}$\times$\color{black}   \\ 
$\Phi=\frac{\pi}{2}$ & $\vartheta =\vartheta(y)$ & $\checkmark$ & $\checkmark$ & \color{red}$\times$\color{black}  \\ 
$\Phi=\frac{\pi}{2}$ & $\vartheta =\vartheta(x,y)$ & $\checkmark$ & $\times$ & \color{red}$\times$\color{black}   \\ 
$0\neq \Phi=\rm{const.} \neq \frac{\pi}{2}$ & $\vartheta =\rm{const.}$ & $\checkmark$ & $\checkmark$ & \color{red}$\times$\color{black}   \\ 
$0\neq \Phi=\rm{const.} \neq \frac{\pi}{2}$ & $\vartheta =\vartheta(x)$ & $\times$ & $\times$ &  \color{green}$\checkmark$\color{black}  \\ 
$0\neq \Phi=\rm{const.} \neq \frac{\pi}{2}$ & $\vartheta =\vartheta(y)$ & $\times$ & $\checkmark$ & \color{red}$\times$\color{black}   \\ 
$0\neq \Phi=\rm{const.} \neq \frac{\pi}{2}$ & $\vartheta =\vartheta(x,y)$ & $\times$ & $\times$ & \color{green}$\checkmark$\color{black}  \\ 
$\Phi=\Phi(x)$ & $\vartheta =\rm{const.}$ & $\checkmark$ & $\times$ & \color{red}$\times$\color{black}   \\ 
$\Phi=\Phi(x)$ & $\vartheta =\vartheta(x)$ & $\times$ & $\times$ &\color{green}$\checkmark$\color{black}   \\ 
$\Phi=\Phi(x)$ & $\vartheta =\vartheta(y)$ & $\times$ & $\times$ & \color{green}$\checkmark$\color{black}   \\ 
$\Phi=\Phi(x)$ & $\vartheta =\vartheta(x,y)$ & $\times$ & $\times$ & \color{green}$\checkmark$\color{black}  \\ 
$\Phi=\Phi(y)$ & $\vartheta =\rm{const.}$ & $\checkmark$ & $\checkmark$ & \color{red}$\times$\color{black}  \\ 
$\Phi=\Phi(y)$ & $\vartheta =\vartheta(x)$ & $\times$ & $\times$ & \color{green}$\checkmark$\color{black}  \\ 
$\Phi=\Phi(y)$ & $\vartheta =\vartheta(y)$ & $\times$ & $\checkmark$ & \color{red}$\times$\color{black}   \\ 
$\Phi=\Phi(y)$ & $\vartheta =\vartheta(x,y)$ & $\times$ & $\times$ & \color{green}$\checkmark$\color{black}   \\ 
$\Phi=\Phi(x,y)$ & $\vartheta =\rm{const.}$ & $\checkmark$ & $\times$ & \color{red}$\times$\color{black}   \\ 
$\Phi=\Phi(x,y)$ & $\vartheta =\vartheta(x)$ & $\times$ & $\times$ & \color{green}$\checkmark$\color{black}  \\ 
$\Phi=\Phi(x,y)$ & $\vartheta =\vartheta(y)$ & $\times$ & $\times$ & \color{green}$\checkmark$\color{black}   \\ 
$\Phi=\Phi(x,y)$ & $\vartheta =\vartheta(x,y)$ & $\times$ & $\times$ & \color{green}$\checkmark$\color{black}   \\ 
\hline
\hline
\end{tabular}
\caption{ 
\textit{SDE resulting from magnetic textures in a model lacking explicit Rashba SOI:} We classify different textures via their angles $\Phi$  and $\vartheta$ and check which symmetries are present in the corresponding class to predict whether a SDE is possible. 
If the spins lie in a plane, then the symmetry $\mathcal{R}_\pi\mathcal{T}$ forbids a SDE. Similar if the texture is not a function of the $x$-coordinate, then the SDE is suppressed in $x$-direction. Both symmetries have to be broken to allow a SDE. Here, we indicate the presence 
[absence] of the symmetries and the SDE with a checkmark [cross]. \label{Tab:SDE_DueToMagTextures} We note that fine-tuned symmetries, e.g. spatial rotational symmetry, can also result in the absence of an SDE and that, even when symmetry allowed, the magnitude of the SDE is not guaranteed to be large.}
\end{table}

\begin{figure}[!t]
\subfloat{\label{figDiodeEffSkyrNoSOI}\stackinset{l}{-0.00in}{t}{-0.0in}{(a)}{\stackinset{l}{1.8in}{t}{-0.05in}{(b)}{\includegraphics[width=1\columnwidth]{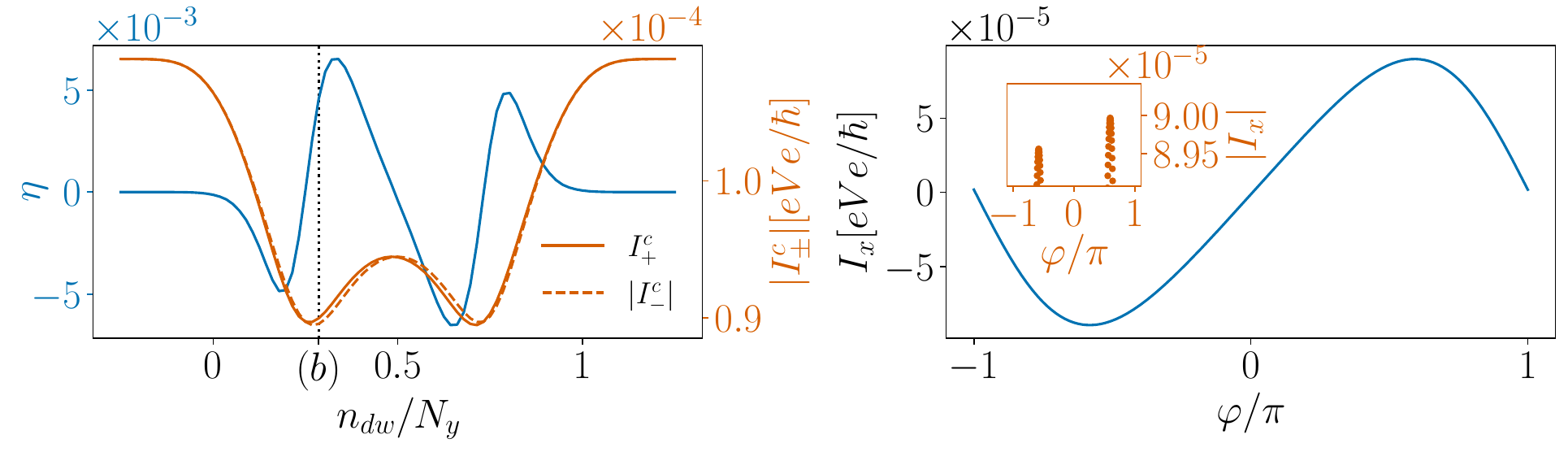}}}
}
\subfloat{\label{figCPRNoSOI}}
\caption{ \textit{ Diode efficiency as a function of the position of the Néel skyrmion in a system without Rashba SOI:} (a) Diode efficiency $\eta$ as a function of the position of the skyrmion $n_{dw}$. (b) The CPR and the modulus of the current (shown in the inset)  indicate the different critical currents. The CPR is calculated for the skyrmion  position as indicated in panel (a). In the inset we plot the data points and do not connect them to show that the resolution of the superconducting phase is indeed high enough to resolve the difference in the critical currents. 
 Parameters: $L_y=2L_x=140$ nm, $m_{\text{eff}}=0.023m_e$, $\mu=5$ meV, $a=2.5$ nm, $\Delta=1$ meV, $J=1$ meV, $\underline{\alpha=0 }$ eVnm, $\gamma=32$ meV, and $\lambda_{s}a=35$ nm. }
 \label{figSDE_Skyrmion_No_SOI}
\end{figure}

To provide a concrete example of a spin texture where an SDE is allowed, we consider skyrmionic textures. In this case both the angles $\vartheta_{\vec r}$ and $\Phi_{\vec r}$ are a function of $x$ and $y$-coordinates and therefore the texture can in principle support a SDE, see Fig.~\ref{figSDE_Skyrmion_No_SOI} where we consider a Néel skyrmion. However, the position of the skyrmion plays a crucial role and the SDE can vanish at certain symmetry points, for example if the skyrmion is placed exactly in the middle of the junction, see Fig.~\ref{figDiodeEffSkyrNoSOI}. 
Comparing the Néel and Bloch type skyrmion we find that the polar angle $\Phi$ agrees for both textures and the azimuthal angle $\vartheta$ can be mapped from the Néel skyrmion to the Bloch skyrmion via the shift $\vartheta\rightarrow \vartheta -\pi/2$. The effective SOI, in Eq.~\eqref{eq:TransformedHamEffSOIAndUniformField}, however, depends only on the derivative of $\vartheta$, but not on the actual value. Consequently, the effective SOI is the same for both configurations, which we also checked numerically. This is also evident from unitary spin rotations around the $z$ axis of the original Hamiltonain $\mathcal{H}$ presented in Eq.~\eqref{Eq:MagneticTextureContinuumModel}, which leave the energy spectrum unchanged. 
 If we add Rashba SOI to the system, then the gauge transformation acting on this additional Rashba term yields an explicit dependence of the transformed Hamiltonian on the angle $\vartheta$, not only on its derivative, therefore the SDE effect differs for the two types of skyrmions if Rashba SOI is present, as was shown in Fig.~\ref{figSkyrmionsAndDiode}.
 
Finally we note that the transformation we utilize to map a varying magnetic texture to an effective SOI is quite general, however, we require that the derivatives of the angles are well-defined, such that the angles should be smooth functions of the coordinates $x$ and $y$. Therefore, we explicitly do not consider certain classes of magnetic textures such as domain walls with $\lambda_{dw}\rightarrow 0$ or antiferromagnetic structures.

\section{Discussion\label{Sec:Conclusion}}
We have shown how a highly controllable SDE can be achieved in a Josephson junction where the normal section is a racetrack hosting magnetic textures, such as domain walls or skyrmions. In particular, the positions of the magnetic texture alters the efficiency of the SDE and can even change its sign, enabling a Josephson transistor effect. First, we showed that a system containing Rashba-like spin-orbit interaction enables a high degree of control that can be exerted on the SDE, depending on the location of the magnetic textures. The ratio between the size of a magnetic texture and the dimensions of the Josephson junction plays an important role in determining the maximal strength of the SDE.  We also showed that certain textures, such as skyrmions, can enable an SDE even in the absence of Rashba SOI in the itinerant charge carrier material and classified some magnetic textures where this is possible.

Our results show that the interplay between magnetic textures and the SDE is an exciting playground for future low-temperature electronics. For instance, this effect could be used to create a superconducting transistor that is controlled by magnetic textures rather than gates. Furthermore, the effects discussed here could be the basis for a low-temperature readout scheme of racetrack memory devices that can be used as components in cryogenic or quantum computers.

\section{Acknowledgements}
We thank Joel Hutchinson and Maximilian Hünenberger for useful conversations. 
This project has received funding from the European Union's Horizon 2020 research and innovation programme under Grant Agreement No 862046 and under  Grant Agreement No 757725 (the ERC Starting Grant). This work was supported by the Georg H. Endress Foundation and the Swiss National Science Foundation. During the preparation of this paper, Ref.~\cite{sinner2023superconducting} appeared on arXiv also proposing the idea that complex magnetic textures in the absence of an explicit Rashba SOI can be sufficient to mediate an SDE.

\appendix
\begin{widetext}
\section{Gauge Transformation}\label{Sec:AppGaugeTrafo}
\subsection{Derivation of the rotated Hamiltonian}
In this Appendix, we present the detailed derivation of the Hamiltonian in Eq.~\eqref{eq:TransformedHamEffSOIAndUniformField}.  Starting point is the Hamiltonian from Eq.~\eqref{Eq:MagneticTextureContinuumModel}, which we split into the kinetic term
\begin{align}
H_0 &= - \frac{\hbar^2}{2m}(\nabla_x^2+\nabla_y^2)\sigma_0
\end{align}
 and the exchange term 
\begin{align}
H_{Ex}&=J \left\lbrace\sin(\Phi)\left[\cos(\vartheta)\sigma_x+\sin(\vartheta)\sigma_y\right]+\cos(\Phi)\sigma_z\right\rbrace.
\end{align}
These terms  transform under the unitary gauge transformation $U_1=e^{-i(\vartheta/2-\pi/4)\sigma_z}$ as
\begin{align}
U_1^{\dagger} H_{Ex}U_1=J\left[\sin(\Phi)\sigma_y+\cos(\Phi)\sigma_z\right],
\end{align}
and
\begin{align}
&U_1^{\dagger}[\nabla_{x_j}^2 \sigma_0]U_1=\left[\nabla_{x_j}^2-\frac{1}{4}\left(\pdv{\vartheta}{x_j}\right)^2 \right]\sigma_0 -i\left[\pdv{\vartheta}{x_j} \nabla_{x_j}+ \frac{1}{2}\pdv[2]{\vartheta}{x_j,}\right]\sigma_z,
\end{align}
where $x_j \in\lbrace x,y\rbrace$, respectively.  Combining the results yields
\begin{align}
U_1^{\dagger}\mathcal{H}U_1=J\left[\sin(\Phi)\sigma_y+\cos(\Phi)\sigma_z\right]-\frac{\hbar^2}{2m}\sum_{x_j\in\lbrace x,y\rbrace} \left\lbrace\left[\nabla_{x_j}^2-\frac{1}{4}\left(\pdv{\vartheta}{{x_j}}\right)^2 \right]\sigma_0 -i\left[\pdv{\vartheta}{{x_j}} \nabla_{x_j}+ \frac{1}{2}\pdv[2]{\vartheta}{{x_j}}\right]\sigma_z\right\rbrace. \label{eq:FirstTrafoApplied}
\end{align}
Next, we apply the second gauge transformation $U_2=e^{i(\Phi/2-\pi/4)\sigma_x}$ on the exchange coupling term like
\begin{align}
U_2^{\dagger}\left[\sin(\vartheta)\sigma_y+\cos(\vartheta)\sigma_z\right]U_2=\sigma_y
\end{align}
to map the system on a ferromagnet with a magnetization in $y$-direction.
The last term in Eq.~\eqref{eq:FirstTrafoApplied} transforms as
\begin{align}
U_2^{\dagger}\sigma_z U_2 =\sin(\Phi)\sigma_z+\cos(\Phi)\sigma_y, \label{eq:SOITerm1}
\end{align}
 while the first derivative term takes the form
\begin{align}
U_2^{\dagger}\nabla_x\sigma_z U_2=&\frac{1}{2}\pdv{\Phi}{x}\left[\cos(\Phi)\sigma_z-\sin(\Phi)\sigma_y\right]+\left[\sin(\Phi)\sigma_z+\cos(\Phi)\sigma_y\right]\nabla_x \label{eq:SOITerm2}.
\end{align}
The combination of the results presented in Eqs.~\eqref{eq:SOITerm1} and \eqref{eq:SOITerm2} allows us  to rewrite those terms as a position dependent SOI, please note the symmetrized form which ensures the hermiticity of the term \cite{Klinovaja2015Fermionic}  
\begin{align}
 &\pdv{\vartheta}{x_j} \left[U_2^{\dagger}\sigma_z\left( \nabla_{x_j}U_2\right)+U_2^{\dagger}\sigma_z U_2\nabla_{x_j}\right]+\frac{1}{2}\pdv[2]{\vartheta}{x_j} U_2^{\dagger}\sigma_z U_2 \nonumber \\
 =& \nabla_{x_j}\left[\frac{1}{2}\pdv{\vartheta}{x_j} \left(\sin(\Phi)\sigma_z+\cos(\Phi)\sigma_y\right)\right]+\left[\frac{1}{2}\pdv{\vartheta}{x_j} \left(\sin(\Phi)\sigma_z+\cos(\Phi)\sigma_y\right)\right]\nabla_{x_j}. 
\end{align}
The second derivative takes the form
\begin{align}
U_2^{\dagger}\nabla_{x_j}^2\sigma_0 U_2
=\nabla_{x_j}^2-\frac{1}{4}\left[\pdv{\Phi}{{x_j}}\right]^2+i\frac{1}{2}\left(\nabla_{x_j}\pdv{\Phi}{{x_j}}+\pdv{\Phi}{{x_j}}\nabla_{x_j}\right)\sigma_x.
\end{align}
Finally, the Hamiltonian is given by 
\begin{align}
U_2^{\dagger}U_1^{\dagger}\mathcal{H}U_1U_2=-\frac{\hbar^2}{2m}\sum_{x_j\in\lbrace x,y\rbrace} \Bigg\lbrace   
\left(\pdv[2]{x_j}-\frac{1}{4}\left[\pdv{\Phi}{x_j}\right]^2-\frac{1}{4}\left[\pdv{\vartheta}{x_j}\right]^2\right)\sigma_0-i \left[\pdv{x_j}\Lambda_{x_j}(x,y)+\Lambda_{x_j}(x,y)\pdv{x_j}\right]
\Bigg\rbrace+J\sigma_y \label{eq:TransformedHamiltonian}
\end{align}
with 
\begin{align}
\Lambda_{x_j}(x,y)=\frac{1}{2}\pdv{\vartheta}{x_j} \left[\sin(\Phi)\sigma_z+\cos(\Phi)\sigma_y\right]-\frac{1}{2}\pdv{\Phi}{x_j}\sigma_x. \label{eq:TransformedHamiltonian2}
\end{align}
Last, we note that the strength of the appearing SOI term does not depend on the strength of the exchange coupling, instead it is only a function of the angles $\vartheta$ and $\Phi$ or their derivatives with respect to the $x$- or $y$-coordinate.

\section{Magnetic textures and the SDE} \label{Sec:AppMagTexAndSDE}
In Sec.~\ref{Sec_SDE_Without_ExplicitSOI}, we discuss which magnetic textures support a SDE, here we numerically study the underlying  conditions and confirm the analytic results. First, we consider a texture that changes in $x$-direction since the angle $\vartheta=g(n_x)$ depends explicitly on the $x$-coordinate. 
Here, the function $g(n_{x_j})=n_{x_j}\pi/\lambda_{m}$ with $n_{x_j}=n_x$ or $n_{x_j}=n_y$ is linear for simplicity and the length-scale $\lambda_{m}$, measured in lattice sites, sets the rotation-period of the magnetic texture. Our choice $\Phi=\frac{\pi}{3}$ forces  the spins into a conical rotation, so that the texture is not confined to a plane. From our analytic analysis we expect a SDE is in principle possible for this case. Indeed, the current-phase relation and in particular the absolute value of the current reveals a finite SDE in $x$-direction, see Fig.~\ref{figMagTexturesAndSDE}. In contrast, if $\vartheta(n_y)=g(n_y)$ varies instead in $y$-direction, then the SDE is suppressed, since the current does not experience any non-uniformity in $x$-direction. Last, we prepare a texture lying in a plane, with $\Phi=g(n_x)$ and $\vartheta=\frac{\pi}{3}$, see the last column in Fig.~\ref{figMagTexturesAndSDE}, the corresponding current is odd-symmetric with respect to $\varphi$ and does therefore not support a SDE as predicted.

\begin{figure*}[t]
\subfloat{\label{figMagTex1}\includegraphics[width=1\textwidth]{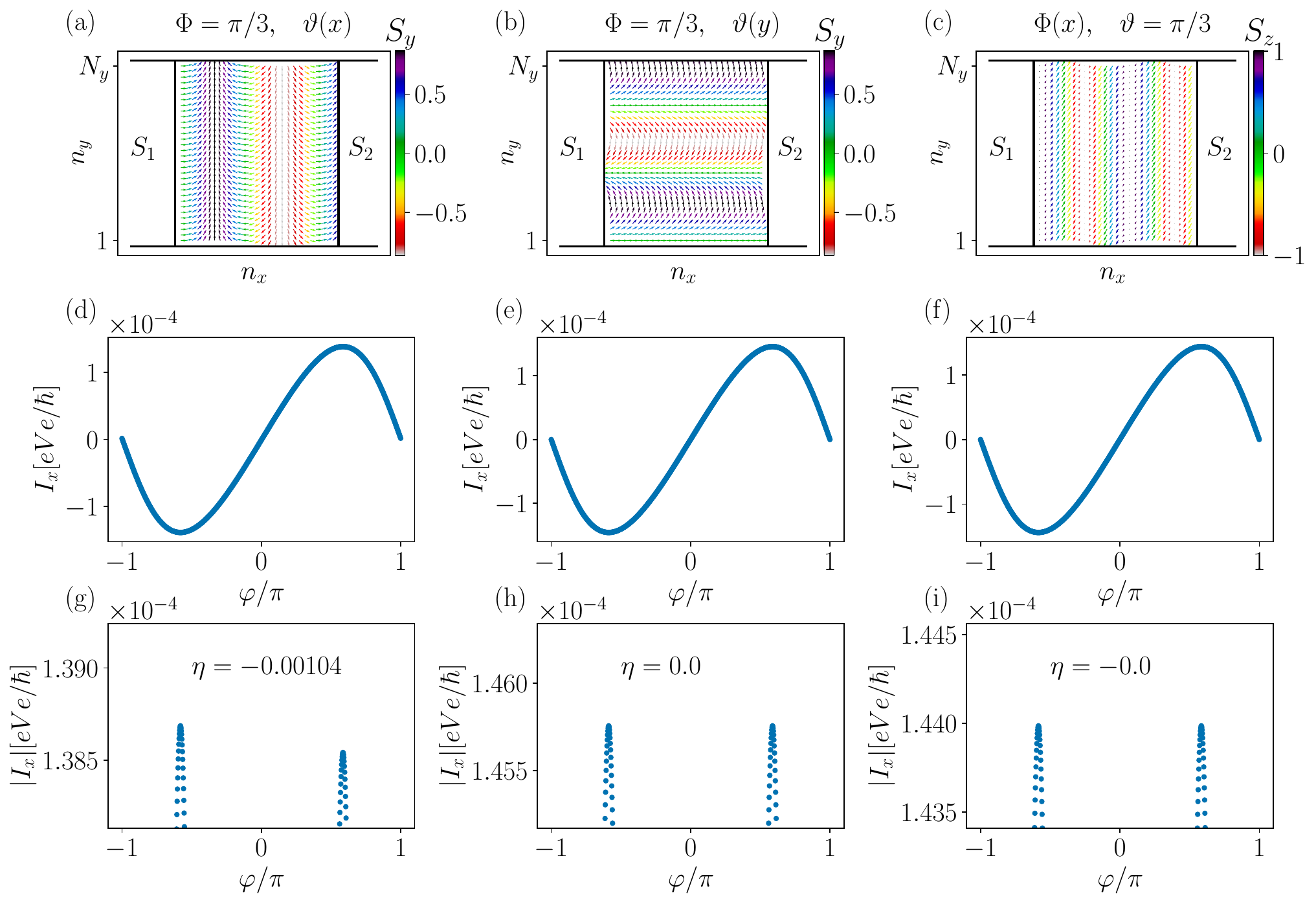}
}
\caption{ \textit{Magnetic textures and the SDE without Rashba SOI}: First row: magnetic textures (a)[b] conical rotation parallel [perpendicular] to the direction of current flow and (c) magnetic texture confined to a plane.  Second row: current phase relations corresponding to the magnetic textures. Third row: absolute value of the current as function of the superconducting phase and associated diode efficiency. The first texture yields a SDE, while the others do not, this is in agreement with our analytic analysis. Parameters: $(L_x, L_y)=(70,80)$ nm, $m_{\text{eff}}=0.023m_e$, $\mu=5$ meV, $a=2.5$ nm, $\Delta=1$ meV, $J=1$ meV, ${\alpha=0} $, $\lambda_{m}a=20$ nm, and $\gamma=32$ meV.}
 \label{figMagTexturesAndSDE}
\end{figure*}

\section{Position of domain walls and skyrmions \label{App:PosDomainWallSkyrmion}}
In this section, we clarify the meaning of negative values of the domain wall or skyrmion centre position, see for example Fig.~\ref{figDomainWallsAndDiode}. We emphasize that the junction has only $N_y$ sites in $y$-direction but we move the domain wall or skyrmion so that only a finite part of it enters the junction, as illustrated in Fig.~\ref{figSkymrionPositions1}, which shows the in-plane magnetization of three system configurations hosting skyrmions: In the first configuration we chose $n_{y,s}=-\lambda_s/2$ so that only a small part of the skymrion is inside the junction. For $n_{y,s}=0$, only the half of a skyrmion has entered the junction. Last, the case $n_{y,s}=\lambda_s$ describes the scenario in which the whole skyrmion just entered the junction. Finally we note that the same logic applies also to the case $n_{dw}>N_y$ [$n_{y,s}>N_y$]. We note that the definitions of the position of the domain wall and the skyrmion are different: 
While the skyrmion position $n_{y,s}$ is measured from the center of the skyrmion, the domain wall position is measured from the beginning of the domain wall, see the definitions in Eqs.~\eqref{eq:DWAngleInPLaneMagnetization} and \eqref{eq:DefSkyrmionTheta}.

 \begin{figure*}[t]
\subfloat{\label{figSkymrionPositions}\stackinset{l}{-0.00in}{t}{0.1in}{(a)}{\stackinset{l}{2.4in}{t}{0.1in}{(b)}{\stackinset{l}{4.8in}{t}{0.1in}{(c)}{\includegraphics[width=1\textwidth]{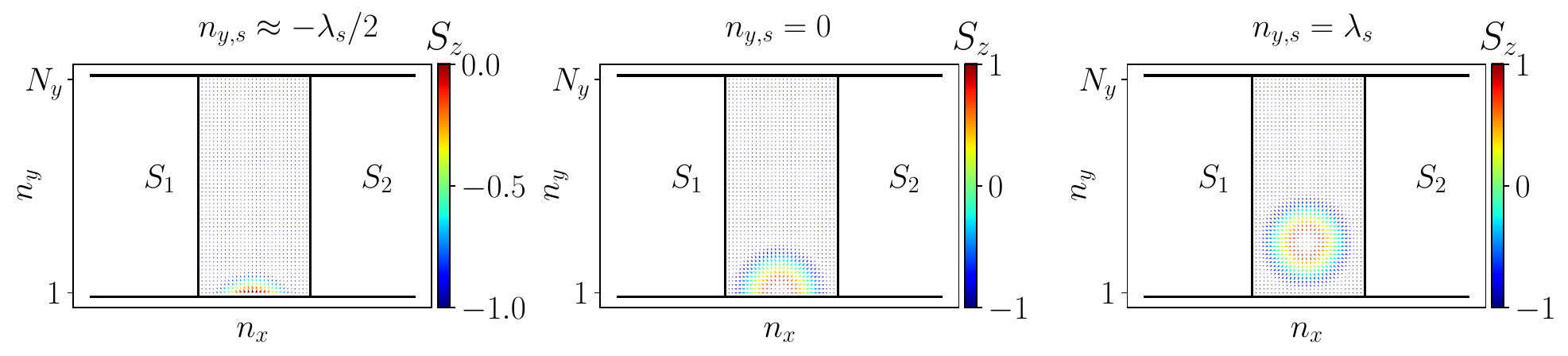}}}}
}
\caption{ \textit{Different positions of a Néel skyrmion}: (a) The skyrmion moved partially into the junction. (b) Half of the skyrmion entered  the junction. (c) The whole skyrmion is fully in the junction. The arrows indicate the in-plane direction of the magnetic moments, while the color-code illustrates the local out of plane magnetization.  The superconducting regions $S_1$ and $S_2$ have with potentially different superconducting phases. The black lines serve as guide lines for the eye to distinguish the normal and superconducting regions and to  highlight the system boundaries.  Parameters: $L_y=2L_x=140$ nm, $a=2.5$ nm, and $\lambda_{s}a=35$ nm.} 
 \label{figSkymrionPositions1}
\end{figure*}
\end{widetext}

\FloatBarrier
\bibliographystyle{apsrev4-1}
\bibliography{Literatur2}
\end{document}